\documentclass[11pt,a4paper]{article}
\pdfoutput=1

\usepackage{amsmath,amssymb,tikz}
\usepackage{graphicx}
 \allowdisplaybreaks
\def\be{\begin{equation}}
\def\ee{\end{equation}}
\def\ba#1\ea{\begin{align}#1\end{align}}
\def\bg#1\eg{\begin{gather}#1\end{gather}}
\def\bm#1\em{\begin{multline}#1\end{multline}}
\def\bmd#1\emd{\begin{multlined}#1\end{multlined}}

\def\({\left(}
\def\){\right)}
\def\[{\left[}
\def\]{\right]}

\def \be {\begin{equation}}
\def \ee {\end{equation}}
\def \ba {\begin{array}}
\def \ea {\end{array}}
\def \bea{\begin{eqnarray}}
\def \eea{\end{eqnarray}}

\def\bea{\begin{eqnarray}}
\def\eea{\end{eqnarray}}

\newcommand{\bit}{\begin{itemize}}  \newcommand{\eit}{\end{itemize}}
\newcommand{\ben}{\begin{enumerate}}  \newcommand{\een}{\end{enumerate}}

\long\def\symbolfootnote[#1]#2{\begingroup%
\def\thefootnote{\fnsymbol{footnote}}\footnote[#1]{#2}\endgroup}

\newcommand{\sysu}{{\it School of Physics and Astronomy, Sun Yat-Sen University, 2 Daxue Road, Zhuhai 519082, China}}

\begin{document}
\thispagestyle{empty}
\begin{center}

~\vspace{20pt}

{\Large\bf Codimension-n Holography for Cones}

\vspace{25pt}

Rong-Xin Miao ${}$\symbolfootnote[1]{Email:~\sf
  miaorx@mail.sysu.edu.cn}

\vspace{10pt}${}$\sysu

\vspace{2cm}

\begin{abstract}
We propose a novel codimension-n holography, called cone holography, between a gravitational theory in $(d+1)$-dimensional conical spacetime and a CFT on the $(d+1-n)$-dimensional defects.  Similar to wedge holography, the cone holography can be obtained by taking the zero-volume limit of holographic defect CFT. Remarkably, it can be regarded as a holographic dual of the edge modes on the defects. For one class of solutions, we prove that the cone holography is equivalent to AdS/CFT, by showing that the classical gravitational action and thus the CFT partition function in large N limit are the same for the two theories. In general, cone holography and AdS/CFT are different due to the infinite towers of massive Kaluza-Klein modes on the branes. We test cone holography by studying Weyl anomaly, Entanglement/R\'enyi entropy and correlation functions, and find good agreements between the holographic and the CFT results. In particular, the c-theorem is obeyed by cone holography. These are strong supports for our proposal.  We discuss two kinds of boundary conditions, the mixed boundary condition and Neumann boundary condition, and find that they both define a consistent theory of cone holography. We also analyze the mass spectrum on the brane and find that the larger the tension is, the more continuous the mass spectrum is. The cone holography can be regarded as a generalization of the wedge holography, and it is closely related to the defect CFT, entanglement/R\'enyi entropy and AdS/BCFT(dCFT). Thus it is expected to have a wide range of applications.
\end{abstract}

\end{center}

\newpage
\setcounter{footnote}{0}
\setcounter{page}{1}

\tableofcontents

\section{Introduction}

The AdS/CFT correspondence plays an important role in our modern understanding of quantum gravity \cite{Maldacena:1997re,Gubser:1998bc,Witten:1998qj}.  As an exact realization of the holographic principle \cite{tHooft:1993dmi,Susskind:1994vu}, it proposes that the quantum gravity theory in an asymptotically anti-de Sitter space (AdS) is dual to the conformal field theory (CFT) on the boundary. Since it is a strong-weak duality, it provides a powerful tool to study the non-perturbative phenomena in gauge theories \cite{Sakai:2004cn,Erlich:2005qh,Sakai:2005yt}, quantum information \cite{Rangamani:2016dms} and condensed matter physics \cite{Hartnoll:2009sz}. 

Many interesting generalizations of AdS/CFT have been developed, which include dS/CFT \cite{Strominger:2001pn,Maldacena:2002vr,Alishahiha:2004md,Alishahiha:2005dj,Dong:2018cuv}, Kerr/CFT \cite{Guica:2008mu,Castro:2010fd}, flat space holography \cite{Bagchi:2010zz,Bagchi:2016bcd}, brane world holography \cite{Randall:1999ee,Randall:1999vf,Karch:2000ct}, surface/state correspondence \cite{Miyaji:2015yva,Takayanagi:2018pml} and AdS/BCFT \cite{Takayanagi:2011zk,Fujita:2011fp,Nozaki:2012qd,Miao:2018qkc,Miao:2017gyt,Chu:2017aab}.  It is remarkable that, in the past few years a doubly holographic model has been proposed for the resolution of information paradox, where the island plays an important role in recovering Page curve of Hawking radiation \cite{Penington:2019npb,Almheiri:2019psf,Almheiri:2019hni}.   See also \cite{Rozali:2019day,Chen:2019uhq,Almheiri:2019psy,Kusuki:2019hcg,Balasubramanian:2020hfs,Sully:2020pza,Geng:2020qvw,Chen:2020uac,Dong:2020uxp,Arias:2019zug,Arias:2019pzy,Geng:2020kxh,Ling:2020laa,Geng:2020fxl,Kawabata:2021hac,Bhattacharya:2021jrn,Kawabata:2021vyo,Geng:2021hlu,Krishnan:2020fer,Deng:2020ent,Chu:2021gdb,Neuenfeld:2021wbl,Neuenfeld:2021bsb,Chen:2020hmv} for related topics.

\begin{figure}[t]
\centering
\includegraphics[width=6.3cm]{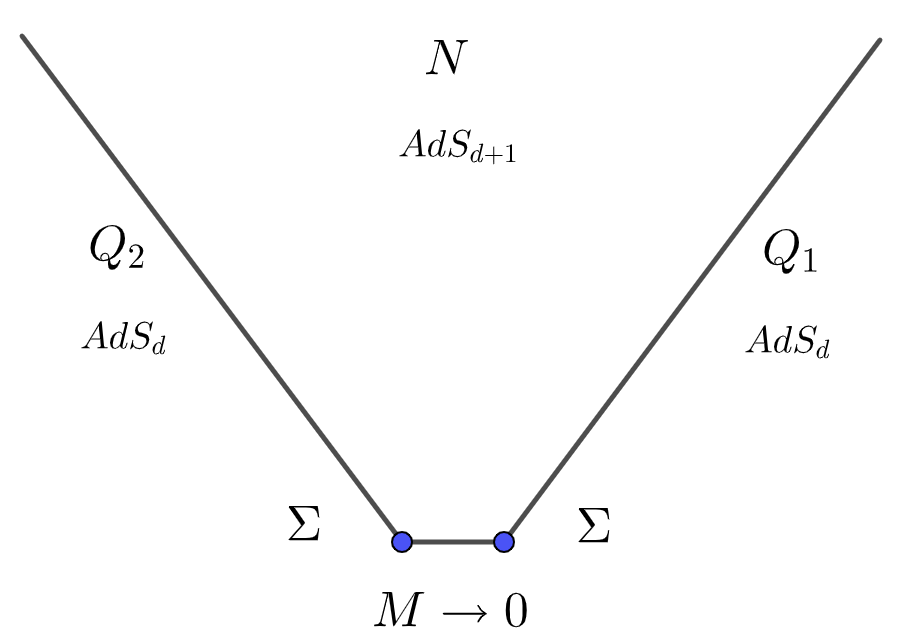}
\includegraphics[width=6cm]{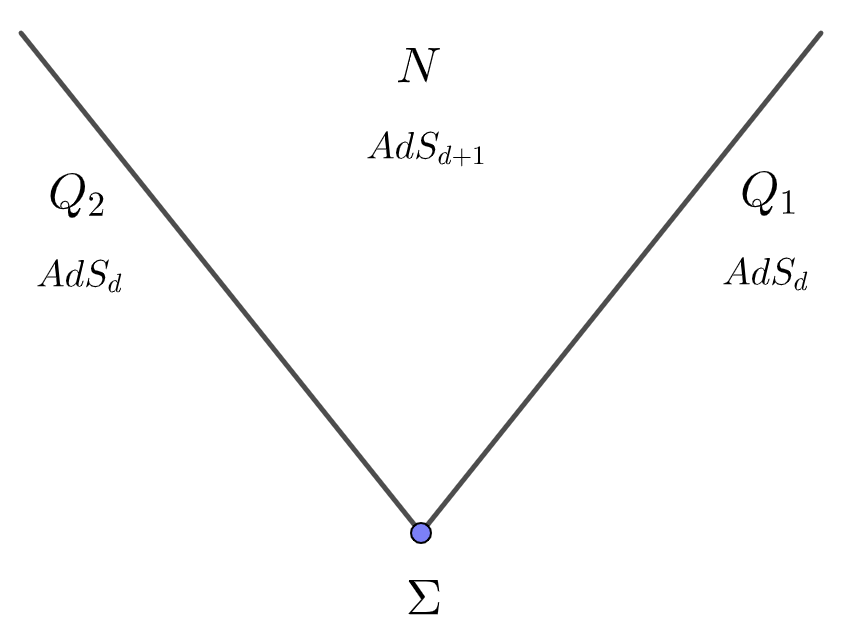}
\caption{ (left) Wedge holography from AdS/BCFT;\  \ (right) Geometry of wedge holography.}
\label{Figurewedge}
\end{figure}

Recently,  a codimension two holography, named wedge holography, is proposed by \cite{Akal:2020wfl} between the gravitational theory in a $(d+1)$-dimensional wedge spacetime and the $(d-1)$-dimensional CFT on the corner of the wedge:
\begin{eqnarray}\label{Wedgeholography}
\text{Gravity on wedge} \ W_{d+1}  \simeq \text{CFT}_{d-1} \ \text{on}\  \Sigma \nonumber
\end{eqnarray}
The geometry of wedge holography is shown in Fig.\ref{Figurewedge} (right), where $N$ denotes $(d+1)$-dimensional wedge space, $Q_1$ and $Q_2$ denote two $d$-dimensional branes, $\Sigma$ is the corner of the wedge where $ \text{CFT}_{d-1}$ lives. See also \cite{Bousso:2020kmy} for a similar proposal of codimension two holography.  Wedge holography can produce the correct free energy, Weyl anomaly, entanglement/R\'enyi entropy, and correlation functions \cite{Akal:2020wfl,Miao:2020oey}.  For one novel class of solutions, it is proved that wedge holography is equivalent to AdS/CFT \cite{Miao:2020oey}.  These are all strong supports for wedge holography. It is interesting that wedge holography can be obtained as a special limit of AdS/BCFT \cite{Takayanagi:2011zk} with vanishing width of a strip \cite{Akal:2020wfl}. See Fig.\ref{Figurewedge} (left) for example. Here BCFT means a conformal field theory defined on a manifold with a boundary, where suitable boundary conditions are imposed. 

\begin{figure}[t]
\centering
\includegraphics[width=12cm]{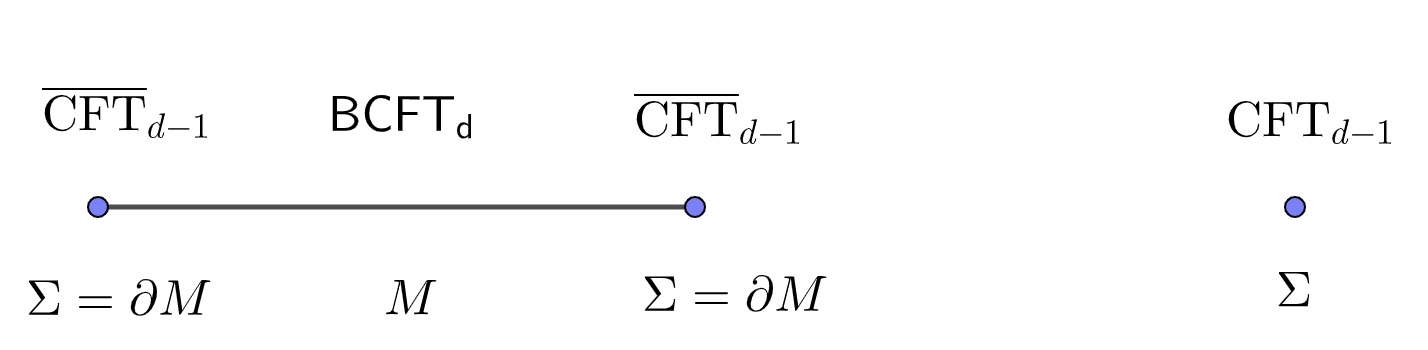}
\caption{(left) $\text{BCFT}_d$ on $M$ and the edge mode as effective $\overline{\text{CFT}}_{d-1}$ on $\Sigma=\partial M$;\  \ (right)  In the zero-volume limit $M\to 0$, $\text{BCFT}_d$ on $M$ disappears and only the edge modes $\text{CFT}_{d-1}=\overline{\text{CFT}}_{d-1}\oplus\overline{\text{CFT}}_{d-1}$ on $\Sigma$ survive.}
\label{FigureCFT2fromBCFT3}
\end{figure}

As a limit of AdS/BCFT, the wedge holography can be regarded as a holographic dual of the edge modes on the boundary.  Let us explain more on this viewpoint. When there is a boundary, in general, there are boundary contributions to Weyl anomaly \cite{Fursaev:2015wpa,Herzog:2015ioa}. Take 3d BCFT as an example, the Weyl anomaly is given by \cite{Jensen:2015swa}
\begin{eqnarray}\label{Weylanomaly3d}
\mathcal{A}=\int_{\partial M} dy^2\sqrt{|\sigma|} \left( b_1 R_{\partial M}+b_2 \text{Tr} \bar{k}^2 \right)
\end{eqnarray}
where $\partial M$, $R_{\partial M}$, $\bar{k}_{ab}$, $b_1, b_2$ denote  the boundary of  manifold $M$,  intrinsic Ricci scalar, traceless parts of extrinsic curvatures and boundary central charges, respectively. 
Remarkably, the first term of (\ref{Weylanomaly3d}) takes the same form as Weyl anomaly of 2d CFTs. Furthermore, the boundary central charge $b_1$ obeys a c-like theorem \cite{Jensen:2015swa}, i.e., $b_{1\ UV}\ge b_{1\ IR}$. This strongly suggests that there are effective CFTs living on the boundary. Note that the effective CFT is a little different from the usual one, since the Weyl anomaly (\ref{Weylanomaly3d}) also depends on the extrinsic curvature $\text{Tr} \bar{k}^2$, which contains the bulk information. Thus it is expected that the holographic dual of such effective CFT is different from the usual one. This novel kind of effective CFT has a natural physical origin, it is the edge mode on the boundary of BCFTs.  Now let us consider the space with two parallel boundaries, such as a strip in Fig.\ref{FigureCFT2fromBCFT3} (left), where the two parallel boundaries are labelled as two blue points. Taking the vanishing volume limit $M\to 0$ so that the two parallel boundaries coincide with each other, the 3d BCFTs living in $M$ disappear and only the edge modes on the boundary $\partial M$ survive. See Fig.\ref{FigureCFT2fromBCFT3} (left) for example, where $d=3$ for our case. In this way, we get effective 2d CFTs from a limit of 3d BCFTs.  Let us go on to discuss the holographic realization of the above approach. According to AdS/BCFT \cite{Takayanagi:2011zk}, the two boundaries ($\Sigma$ of  Fig.\ref{FigureCFT2fromBCFT3} (left)) are extended to two end-of-world branes $Q_1$ and $Q_2$ in the bulk $N$, and the gravity theory in the bulk $N$ is dual to the BCFT on $M$.  See Fig.\ref{Figurewedge} (left) for example. By taking the zero-volume limit of AdS/BCFT, i.e., $M\to 0$,  we are left with the edge mods on $\Sigma$, and we finally obtain the wedge holography as shown in Fig.\ref{Figurewedge} (right), which can be regarded as a holographic dual of edge modes as we have argued above.


Some comments are in orders. {\bf 1}. The dependence on extrinsic curvatures by Weyl anomaly (\ref{Weylanomaly3d}) implies that the edge modes contain the bulk information. As a result, the 2d edge mode is not dual to a 3d gravity as usual. Instead, it is dual to the gravity theory in a 4d wedge spacetime. {\bf 2}. The above discussions can be generalized to general dimensions. For even d, the boundary term of Weyl anomaly does not include intrinsic Euler density. This does not means there are no edge modes on the boundary. In fact, there are always boundary entropy on the boundary, which decreases under RG flow and is a strong evidence for the existence of boundary states/ edge modes.  {\bf 3}.  Let us summarize the steps for the construction of wedge holography as a holographic dual of edge modes. First, study Weyl anomaly to see if the edge modes behave effectively as CFTs on the defect. Second, take suitable zero-volume limit so that only the edge modes survive. Third, extend the discussions into the bulk to obtain a holographic dual of the edge modes. In other words, take suitable limit of AdS/BCFT to get wedge holography. 

So far we focus on the codim-1 defect (boundary). It is interesting to generalize the discussions to general defect CFT (dCFT). This is the main purpose of this paper.  We follow the above steps for the  construction of wedge holography and generalize it to codim-m defects. Let us first study the Weyl anomaly. Without loss of generality, we consider a codim-2 defect in four dimensions. The Weyl anomaly takes the following form 
\begin{eqnarray}\label{Weylanomaly4ddefect1}
\mathcal{A}&=&\int_{M} dx^4\sqrt{|g|} (\frac{c}{16\pi^2} C^{ijkl}C_{ijkl}-\frac{a}{16\pi^2}E_4)\\
&& +\int_{D} dy^2\sqrt{|\sigma|} \left( d_1 R_{D}+d_2C^{ab}_{\ \ ab}+d_3 \text{Tr} \bar{k}^2 \right),\label{Weylanomaly4ddefect2}
\end{eqnarray}
where $C_{ijkl}$, $E_4$, $R_D$, $\bar{k}$ are the Weyl tensor, Euler density in the bulk $M$, intrinsic Ricci scalar and traceless parts of extrinsic curvatures on the defect $D$, respectively.  Here $C^{ab}_{\ \ ab}$ denotes the contraction of the Weyl tensor projected to directions orthogonal to $D$, and $(a,c,d_1,d_2,d_3)$ are central charges. 
Similar to the case of BCFT, the first term of (\ref{Weylanomaly4ddefect2}) takes the same form as Weyl anomaly of 2d CFTs and the corresponding central charge $d_1$ obeys a c-like theorem \cite{Jensen:2015swa}, i.e., $d_{1\ UV}\ge d_{1 \ IR}$.  This means that the edge modes on the defect behave effectively as 2d CFTs.  Since the Weyl anomaly (\ref{Weylanomaly4ddefect2}) on the defect $D$ also depends on the bulk Weyl tensions and extrinsic curvatures, the edge modes contain the bulk information and are different from the usual 2d CFTs.  This is also similar to the case of BCFT.  

\begin{figure}[t]
\centering
\includegraphics[width=9cm]{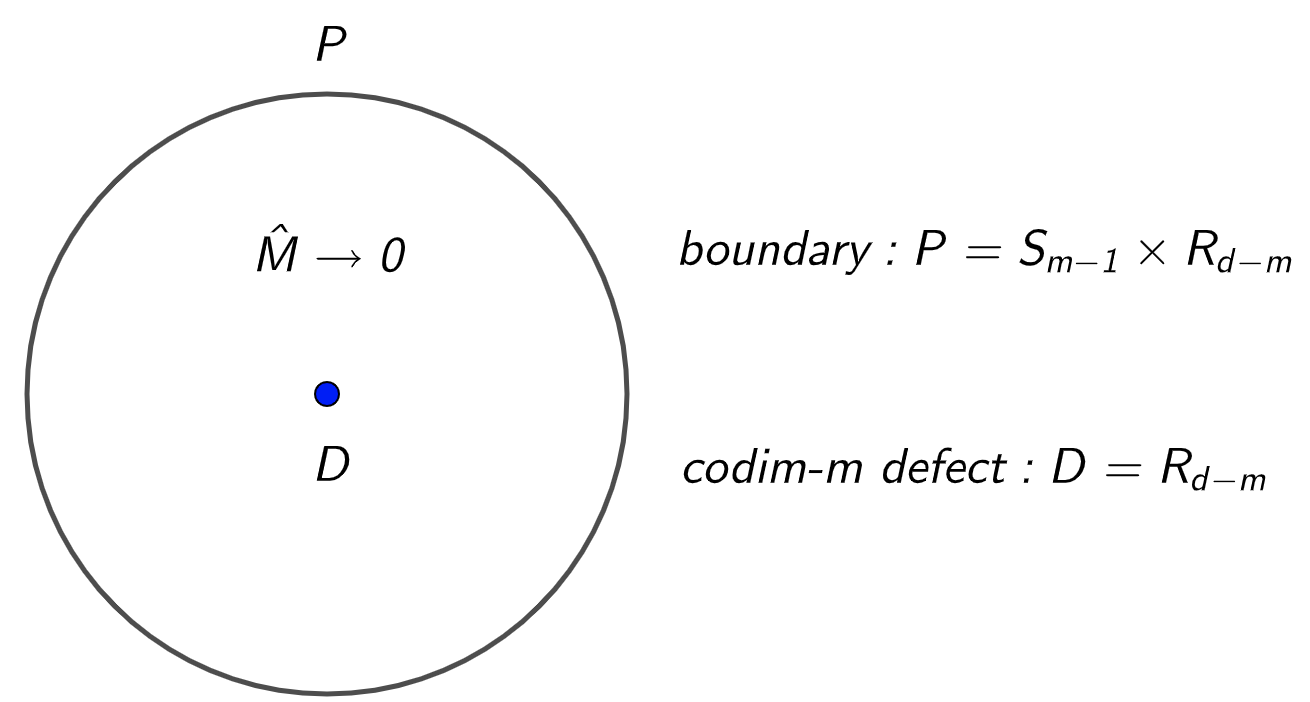}
\caption{Zero-volume limit of dCFT.  $\hat{M}$ is a d-dimensional manifold where dCFT is defined, $P$ is the boundary of $\hat{M}$ and $D$ is a codim-m defect at the center of $\hat{M}$. The metric is given by (\ref{metricdCFT}), $ds^2=dz^2+\frac{z^2}{q^2} d\Omega_{m-1}^2+\sum_{\hat{i}=1}^{d-m} dy_{\hat{i}}^2$ with $0\le z\le z_0$. The defect $D$ and the boundary $P$ are located at $z=0$ and $z=z_0$, respectively. Note that the geometry of $P$ is chosen to be $S_{m-1}\times R_{d-m}$ so that it coincides with the defect $D=R_{d-m}$ in the zero-volume limit $\hat{M}\to 0$ with $z_0\to0$ and $S_{m-1}\to 0$. Here $z_0$ is the radius of the sphere $S_{m-1}$. In such limit, only the edge modes of dCFT survive.}
\label{FiguredCFT}
\end{figure}

Now let us turn to the second step to take a suitable zero-volume limit. To have a well-defined zero-volume limit, we add a  boundary $P$ which surrounds the codim-2 defect D as shown in Fig.\ref{FiguredCFT}.  We require the geometry of $P$ to be $S_{1}\times R_{2}$ so that it coincides with the codim-2 defect $D=R_2$ in the zero-volume limit with $S_1\to 0$.  See Fig. \ref{FiguredCFT} for example, where $d=4$ and $m=2$ for our present case.  Let us explain the above constructions in more details. Consider the following metric for general d and m
\begin{eqnarray}\label{metricdCFT}
ds^2=dz^2+\frac{z^2}{q^2} d\Omega_{m-1}^2+\sum_{\hat{i}=1}^{d-m} dy_{\hat{i}}^2,
\end{eqnarray}
where $q$ is a positive constant related to the conical singularity, $d\Omega_{m-1}^2$ is the line element of the unit sphere, the codim-m defect $D$ is located at $z=0$, the boundary $P$ is at $z=z_0$. From (\ref{metricdCFT}) it is clear that, the boundary $P=S_{m-1}\times R_{d-m}$ coincides with the codim-m defect $D=R_{d-m}$ in the zero-volume limit,
\begin{eqnarray}\label{zerolimitdCFT}
\lim_{z_0\to 0} P\simeq D=R_{d-m},
\end{eqnarray}
where we have used the fact that the sphere $S_{m-1}$ shrinks to zero in the limit of zero radius $z_0\to 0$.  Before we go to the third step, let us discuss more on the edge modes. For simplicity, we return to the case with $d=4$ and $m=2$.  Before we take the limit $z_0\to 0$, there are two kinds of edge modes: one lives on the 2d defect $D$ and the other one lives on 3d boundary $P$.  After we perform the limit $z_0\to 0$, the circle $S_1\to 0$ shrinks to zero.  Due to the Kaluza-Klein mechanism, the 3d edge modes on $P$ become effectively 2d fields, which include massless modes and infinite towers of massive modes. Since the massive modes have infinite mass $\hat{m}_k \sim k/z_0 \to \infty$, they decouple from the massless modes and can be ignored safely at finite energy scale. As a result, the edge modes on the 3d boundary $P$ become massless KK mode on the 2d defect $D\simeq\lim_{z_0\to 0}P$ in the zero-volume limit. Now the two kinds of edge modes both live on the 2d defect $D$ effectively in the zero-volume limit. Clearly, this is also the case for general d and m. 
 \begin{figure}[t]
\centering
\includegraphics[width=12cm]{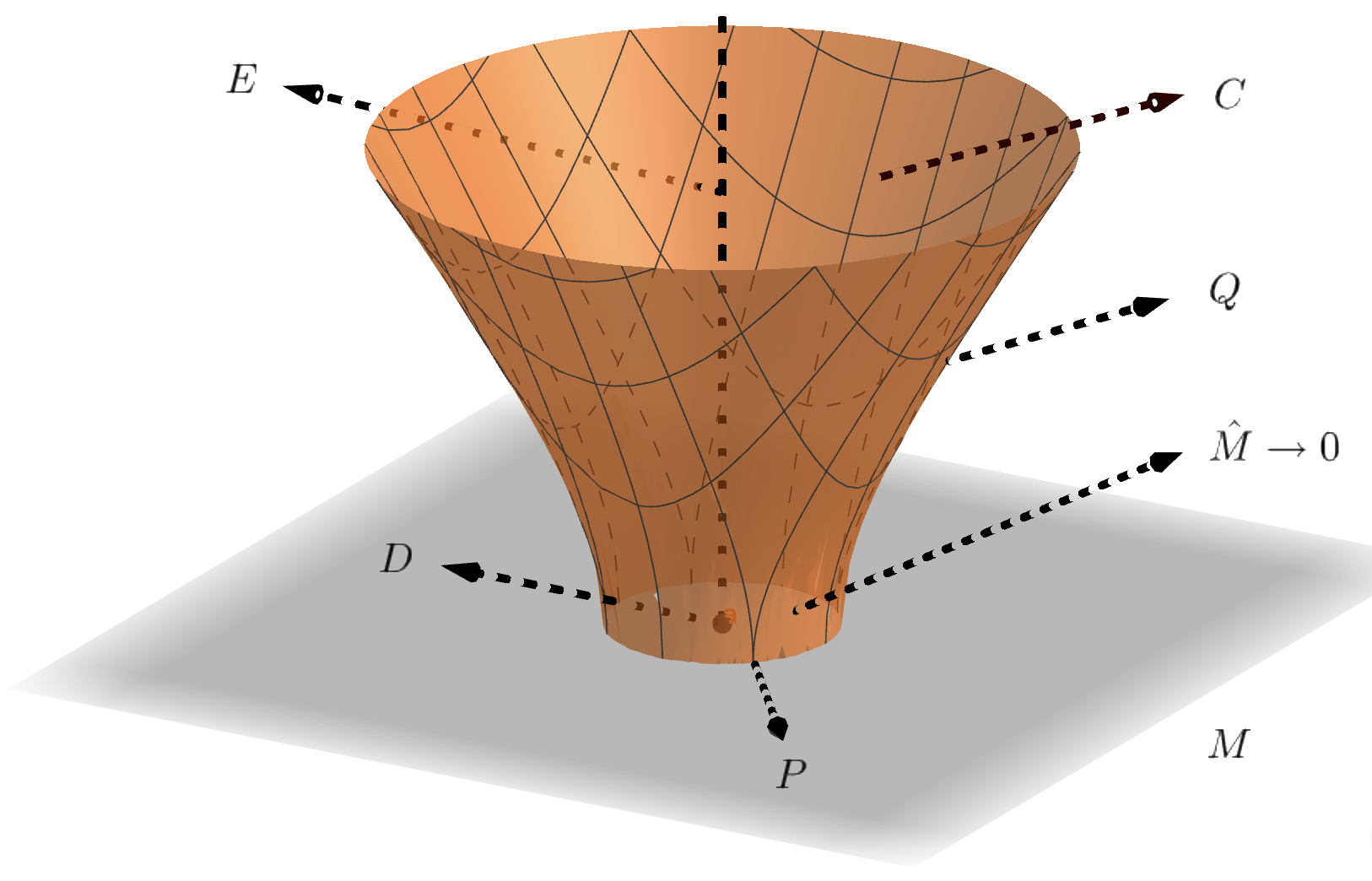}
\caption{Cone holography from AdS/BCFT and AdS/dCFT.  dCFT lives in the manifold $\hat{M}$ with a boundary $P$ and a codim-m defect D at the center. The boundary P and codim-m defect D are extended to an end-of-world brane $Q$ and a codim-m  brane $E$ in the bulk, respectively. $C$ (orange) is the bulk spacetime bounded by $Q$ and $\hat{M}$, $M$ (gray) is the AdS boundary. In the limit $\hat{M}\to 0$, the bulk spacetime $C$ becomes a cone and we obtain the cone holography from AdS/BCFT and AdS/dCFT.}
\label{FigureconeholographyfromAdSBCFT}
\end{figure}

\begin{figure}[t]
\centering
\includegraphics[width=12cm]{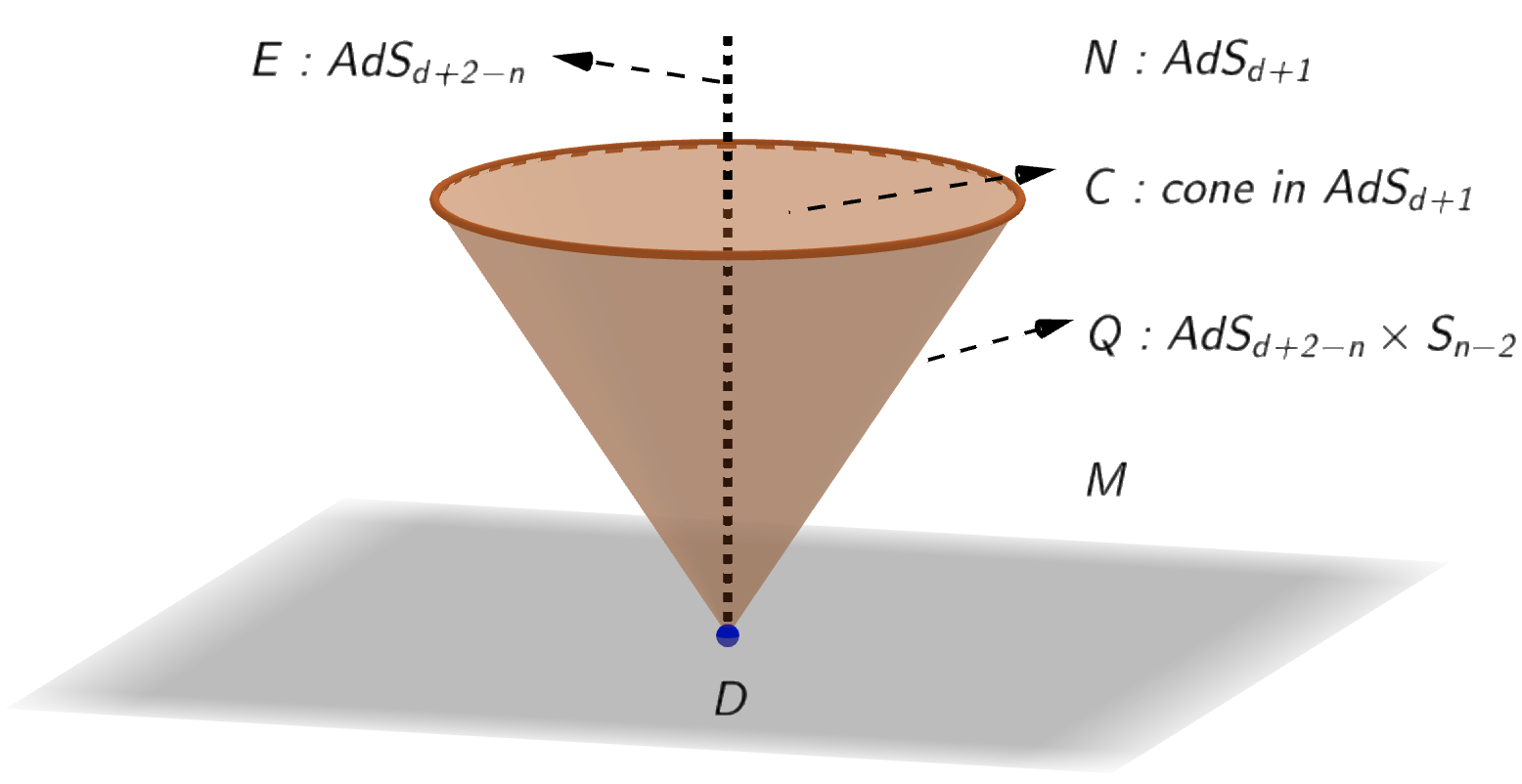}
\caption{ Geometry of cone holography: $M$ is a d-dimensional manifold (gray plane), $D$ is a codim-m defect (blue point) in $M$, where $m=n-1$. $M$ is extended to a (d+1)-dimensional asymptotically AdS space $N$, and $D$ is  extended to a (d+1)-dimensional cone $C$ (orange) in the bulk. The cone $C$ is bounded by a codim-1 brane $Q$ (boundary of orange cone), i.e., $\partial C=Q$. The geometries of $Q$ and $E$ are set to be $\text{AdS}_{d+2-n}\times \text{S}_{n-2}$ and $\text{AdS}_{d+2-n}$ so that they shrink to the same defect $D=\partial Q=\partial E$ on the AdS boundary $M$. The cone holography proposes that a gravity theory in the (d+1)-dimensional cone $C$ is dual to a CFT on (d+1-n)-dimensional defect $D$. }
\label{Figurecodimnholography}
\end{figure}

Now let us go to step 3 to construct a holographic dual of the edge modes on defects.  Recall that dCFT lives in the manifold $\hat{M}$ with a boundary $P$ and a codim-m defect $D$ at the center. See Fig.\ref{FiguredCFT}. We first consider a small but finite $\hat{M}$. Following AdS/BCFT \cite{Takayanagi:2011zk} and AdS/dCFT \cite{Jensen:2013lxa,DeWolfe:2001pq,Dong:2016fnf} 
\footnote{In fact, AdS/BCFT can be regarded as a special case of AdS/dCFT, since the boundary is a codim-1 defect. For our purpose, we want to distinguish the codim-1 defect and the codim-m defect with $m\ge2$. Thus, by `dCFT', we means the codim-m defect with $m\ge 2$ in this paper. }, the boundary $P$ and the codim-m defect $D$ on $\hat{M}$ are extended to an end-of-world brane $Q$ and a codim-m brane $E$ in the bulk, respectively. Please see Fig.\ref{FigureconeholographyfromAdSBCFT} for the geometry, where $C$ is the bulk spacetime bounded by $Q$ and $\hat{M}$, i.e., $\partial C=Q\cup \hat{M}$.  According to AdS/BCFT \cite{Takayanagi:2011zk} and AdS/dCFT \cite{Jensen:2013lxa,DeWolfe:2001pq,Dong:2016fnf}, a gravity theory in the bulk $C$ is dual to the dCFT defined on $\hat{M}$.  Now let us take the zero-volume limit $\hat{M}\to 0$. One the AdS boundary, the boundary $P$ and codim-m defect $D$ coincide and only edge modes of dCFT survive. On the other hand, in the bulk, $C$ becomes a conical spacetime when $\hat{M}\to 0$. Thus, the gravity theory in the (d+1)-dimensional conical spacetime $C$ is dual to the edge modes (CFT) on the (d-m)-dimensional defect $D$. We call this novel holography as cone holography or codim-n holography, where $n=m+1$. See  Fig.\ref{Figurecodimnholography} for the geometry of cone holography, which we will explain more in section 2.  Note that the above arguments can be regarded as a derivation of cone holography from AdS/dCFT following the same logic of the derivation of wedge holography from AdS/BCFT \cite{Akal:2020wfl}.  In the followings of this paper, we provide more evidences for this proposal. 

Let us summarize the main results of this paper. We propose a codim-n holography, called cone holography, between the gravitational theory in a $(d+1)$-dimensional conical spacetime and the CFT on a $(d+1-n)$-dimensional defects.  We discuss two kinds of boundary conditions for the end-of-world brane $Q$, the mixed boundary condition and Neumann boundary condition, and show that they both define a consistent theory. We test our proposal by studying holographic Weyl anomaly, holographic entanglement/R\'enyi entropy, holographic correlation functions and find good agreements with results of CFTs. Besides, we find that the c-theorem is obeyed by cone holography. These are all strong supports for our proposal. Finally, we discuss the mass spectrum of cone holography and find that there are infinite towers of massive gravity on the brane. In the limit of small brane tensions, the massive modes are frozen at low energy and the effective theory on the brane is Einstein gravity. 

The paper is organized as follows. 
In section 2, we formulate the cone holography and prove that it is equivalent to AdS/CFT with Einstein gravity for a novel class of solutions. In section 3, we test cone holography by studying the holographic Weyl anomaly, holographic R\'enyi entropy and correlation functions. In section 4, we discuss carefully the cone holography with Neumann BC. In section 5, we discuss more general solutions to cone holography and show that the gravity on the brane is massive generally. Finally, we conclude with some open problems in section 6. 

Notations: The coordinates on $C$, $Q$, $E$ and $D$ are labeled by $X^{A}=(r, x^{a},y^i)$, $x^{\mu}=(x^{a}, y^i)$, $y^i=(z, y^{\hat{i}})$ and $y^{\hat{i}}$, respectively, where $A$ runs from 1 to $d+1$, $a$ runs from 2 to $n-1$ and $i$ runs from $n$ to $d+1$.  Besides, $g_{AB}, h_{\mu\nu}$, $\gamma_{ij}$ and $\sigma_{\hat{i}\hat{j}}$ denote the metrics on $C$, $Q$, $E$ and $D$. $D$ is a codim-m defect on the AdS boundary $M$, and $E$ is a codim-m brane in the bulk $N$, where $m=n-1$.

\section{Cone holography}

In this section, we formulate the general theory of cone holography. We discuss the geometry, the boundary conditions and solutions. We prove that the gravitational action of cone holography is equivalent  to that of AdS/CFT for one general class of solutions. Assuming that AdS/CFT holds, this can be regarded as a proof of cone holography in a certain sense. For simplicity, we mainly focus on the mixed BC in this section. We leave a careful study of Neumann BC in sect. 4. 

\subsection{Cone holography}

\subsubsection{Geometry}
To start, let us recall the geometry of cone holography as shown in Fig.\ref{FigureconeholographyfromAdSBCFT} and Fig.\ref{Figurecodimnholography}. Let us first illustrate Fig.\ref{FigureconeholographyfromAdSBCFT}. $\hat{M}$ is the d-dimensional manifold where the dCFT is defined, $P$ is the boundary of $\hat{M}$ and $D$ is a codim-m defect at the center of $\hat{M}$. The geometry of the boundary is set to be $P=S_{m-1}\times D$ so that it shrinks to the defect $D$ in the zero-volume limit with $\hat{M}\to 0$ and $S_{m-1}\to 0$. Following AdS/BCFT \cite{Takayanagi:2011zk} and AdS/dCFT \cite{Jensen:2013lxa,DeWolfe:2001pq,Dong:2016fnf}, the boundary $P$ and codim-m defect $D$ on $\hat{M}$ are extended to an end-of-world brane $Q$ and a codim-m brane $E$ in the bulk, respectively. $C$ is the bulk manifold bounded by $Q$ and $\hat{M}$, i.e., $\partial C=Q\cup \hat{M}$. According to AdS/BCFT \cite{Takayanagi:2011zk} and AdS/dCFT \cite{Jensen:2013lxa,DeWolfe:2001pq,Dong:2016fnf}, the gravity theory in the bulk $C$ is dual to the dCFT on $\hat{M}$. In the zero-volume limit $\hat{M}\to 0$, the dCFT on $\hat{M}$ disappears and only the edge modes on the defect $D\simeq \lim_{\hat{M}\to 0} P$ survive. As a result, the gravity theory in the bulk $C$ is dual to the  edge modes on the defect $D\simeq \lim_{\hat{M}\to 0} P$ in the zero-volume limit $\hat{M}\to 0$.  In this way, we derive cone holography from a suitable limit of AdS/dCFT. 

Let us go on to explain Fig.\ref{Figurecodimnholography}, which is obtained from Fig.\ref{FigureconeholographyfromAdSBCFT} by taking the limit $\hat{M}\to 0$. $M$ is a d-dimensional manifold, $D$ is a codim-m defect in $M$, where $m=n-1$.  In AdS/CFT,  $M$ is extended to a (d+1)-dimensional asymptotically AdS space $N$.  In cone holography, the defect $D$ is extended to a (d+1)-dimensional cone $C$ in the bulk, which is bounded by an end-of-world brane $Q$, i.e., $\partial C=Q$. This should be understood as a limit of AdS/dCFT as shown in Fig.\ref{FigureconeholographyfromAdSBCFT}: $P$ is extended to $Q$, $D$ is extended to $E$ and $\hat{M}$ is extended to $C$. Note that the geometries of $Q$ and $E$ are required to be $\text{AdS}_{d+2-n}\times \text{S}_{n-2}$ and $\text{AdS}_{d+2-n}$ so that they shrink to the same defect $D=\partial  Q=\partial E$ on the AdS boundary $M$. This is a key characteristic of cone holography, which enables the theory to be 
codim-n. It should be stressed that, as a solution to Einstein equation, the bulk geometry is smooth everywhere. The conical singularity can only appear on the defect $D$. The cone holography proposes that
\begin{eqnarray}\label{coneholographyproposal}
\text{Classical gravity in the cone} \ C_{d+1}  \simeq \text{CFT}_{d+1-n} \ \text{on the defect } D. \nonumber
\end{eqnarray}
It is a natural generalization of wedge holography \cite{Akal:2020wfl}.  Similar to  wedge holography \cite{Akal:2020wfl}, the cone holography can be obtained as a suitable limit of AdS/BCFT \cite{Takayanagi:2011zk,Fujita:2011fp,Nozaki:2012qd,Miao:2018qkc,Miao:2017gyt,Chu:2017aab} and AdS/dCFT \cite{Jensen:2013lxa,DeWolfe:2001pq,Dong:2016fnf}.  As we have explained in the introduction and in the above paragraph, cone holography can be regarded as a holographic dual of edge modes on defects.  From now on, we forget the origin from edge modes and take cone holography as a general theory of holography. We label the cone holography by $\text{AdSC}_{d+1}/\text{CFT}_{d+1-n}$ in this paper. 

\begin{figure}[t]
\centering
\includegraphics[width=12cm]{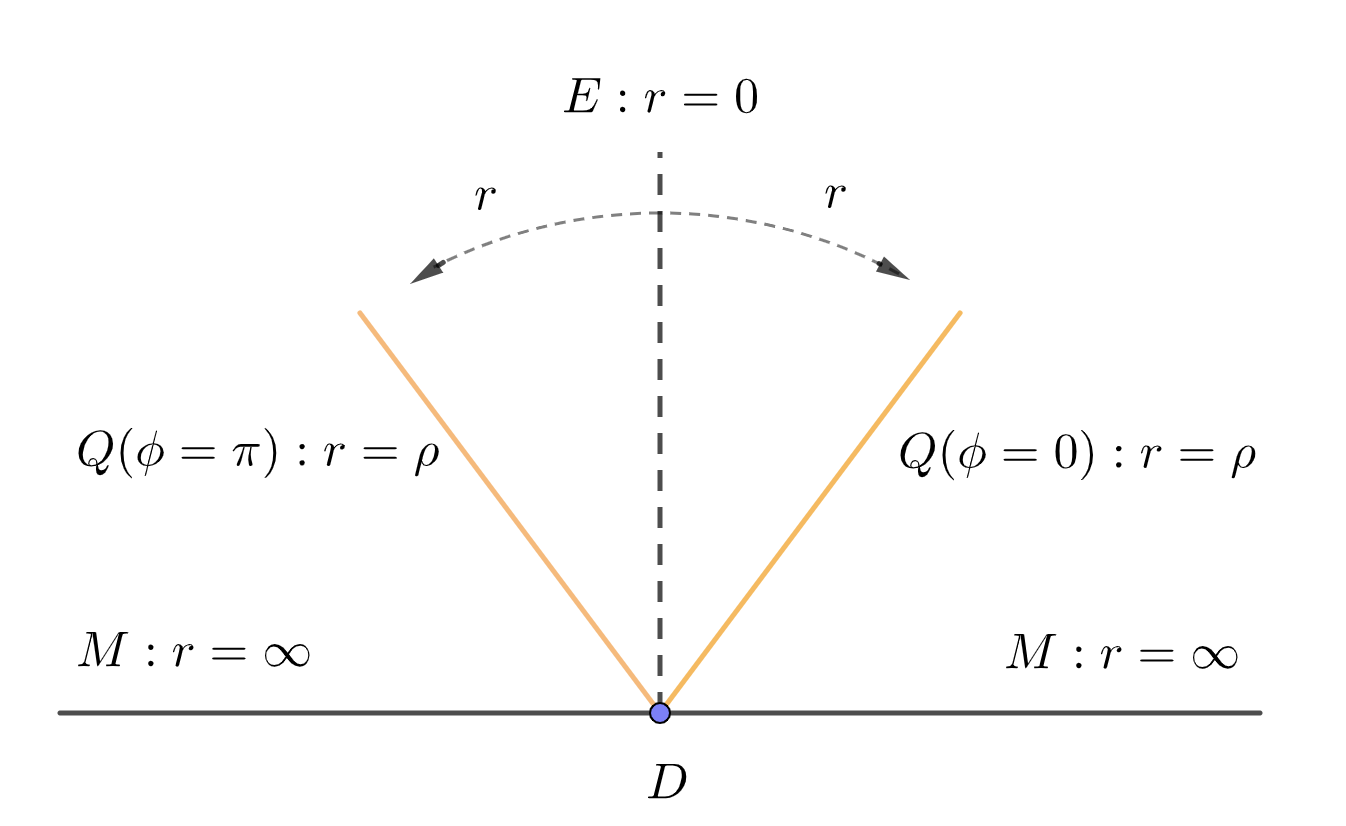}
\caption{ A sketch of cone holography for $\phi=0$ and $\phi=\pi$, where $\phi \simeq \phi+2\pi$. }
\label{Figureconemetric}
\end{figure}

To get a better understanding of the geometry of Fig.\ref{Figurecodimnholography}, let us study a typical metric of cone holography
\begin{eqnarray}\label{conemetric}
ds^2=dr^2+\sinh^2( r) d\Omega_{m-1}^2+\cosh^2 (r) \frac{dz^2+\sum_{\hat{i}=1}^{d-m} dy_{\hat{i}}^2}{z^2}, \ \ \ 0\le r \le \rho,
\end{eqnarray}
which is a locally AdS space. Here $r$ is the proper distance to the codim-m brane $E$, $d\Omega_{m-1}^2$ is the line element of the unit sphere, $m=n-1$ and $\rho$ is a constant. The codim-m brane $E$, the end-of world brane $Q$ and the AdS boundary $M$ are located at $r=0$, $r=\rho$ and $r=\infty$, respectively. The codim-m defect $D$ is at $z=0$ on the AdS boundary $M$.  Please see Fig.\ref{Figureconemetric} for a sketch of the geometry of cone holography for $\phi=0$ and $\phi=\pi$, where $\phi$ is the angle of $S_{m-1}$ with the period $2\pi$. 
From (\ref{conemetric}), we can read off the induced metrics on the branes $E$ and $Q$ as
\begin{eqnarray}\label{conemetricE}
&&ds_E^2= \frac{dz^2+\sum_{\hat{i}=1}^{d-m} dy_{\hat{i}}^2}{z^2}, \\
&&ds_Q^2= \sinh^2(\rho) d\Omega_{m-1}^2+\cosh^2 (\rho) \frac{dz^2+\sum_{\hat{i}=1}^{d-m} dy_{\hat{i}}^2}{z^2}, \label{conemetricQ}
\end{eqnarray}
which clearly shows that the geometries of $E$ and $Q$ are  $\text{AdS}_{d+2-n}$ and $\text{AdS}_{d+2-n}\times \text{S}_{n-2}$, as it has been shown in Fig.\ref{Figurecodimnholography} (recall that n=m+1). Let us go on to discuss the induced metric on the AdS boundary $M$ ($r=\infty$). From (\ref{conemetric}), we have 
\begin{eqnarray}\label{conemetricM}
ds_M^2\sim  \lim_{r\to \infty} \frac{e^{2r}}{z^2}\left( z^2 d\Omega_{m-1}^2+dz^2+\sum_{\hat{i}=1}^{d-m} dy_{\hat{i}}^2 \right),
\end{eqnarray}
which is conformally equivalent to the metric of dCFT (\ref{metricdCFT}) with $q=1$. Recall that $q$ is a parameter related to the conical singularity, which is defined in (\ref{metricdCFT}). When $q=1$, there is no conical singularity on the defect $D$, i.e., $z=0$.  To discuss more general defects with conical singularities, one can replace the metric (\ref{conemetric}) by
\begin{eqnarray}\label{conemetricgeneralfg}
ds^2=dr^2+f(r) d\Omega_{m-1}^2+g(r) \frac{dz^2+\sum_{\hat{i}=1}^{d-m} dy_{\hat{i}}^2}{z^2},
\end{eqnarray}
where $f(r)$ and $g(r)$ can be determined by solving Einstein equations with suitable boundary conditions. We will discuss this more general metric later.


\subsubsection{Action and boundary conditions}
The gravitational action of cone holography is given by
\begin{eqnarray}\label{action}
  I=\frac{1}{16\pi G_N}\int_C  dX^{d+1}\sqrt{|g|} (R-2\Lambda)
  +\frac{1}{8\pi G_N}\int_{Q} dx^d\sqrt{|h|} (K-T)-T_E \int_E  dy^{d+1-m}\sqrt{|\gamma|},
\end{eqnarray}
where $R$ is the Ricci scalar, $-2\Lambda=d(d-1)$ is the cosmological constant (we have set AdS radius $L=1$), $K$ is the extrinsic curvature, $T$ and $T_E$ are the tensions of branes $Q$ and $E$. $g_{AB}, h_{\mu\nu}, \gamma_{ij}$ are metrics on $C, Q, E$, respectively.  Note that there is no well-defined
“thin brane” limit for the Einstein equations for codimension 3 and higher unless the tensions $T_E$ vanish \cite{Geroch:1986jjl,Charmousis:2001hg,Gherghetta:2000jf,Bostock:2003cv}. Thus we have
\begin{eqnarray}\label{braneETE}
 T_E\to 0, \ \ \text{for } m\ge 3.
\end{eqnarray}
As a result, the last term of action (\ref{action}) should be understood as a probe brane with $T_E\to 0$ for $m\ge 3$, which is added to help us to determine the location of branes.  To have non-zero tensions for $m\ge 3$, one can consider either higher derivative gravity or thick branes \cite{Bostock:2003cv}.  On the other hand, the codim-2 brane ($m=2$) can have non-zero tension $T_E$. And the tension is related to the conical singularity by $T_E\sim (q-1)/q$ \cite{Dong:2016fnf}. 

One of the central tasks of cone holography is to determine the locations of the two kinds of branes ($E$ and $Q$).  Let us first discuss the codim-m brane $E$. Inspired by Ryu-Takayanagi formula \cite{Ryu:2006bv}, we propose that $E$ is a minimal surface in the bulk whose boundary is given by the defect $D$, i.e., $\partial E=D$.  One trick to derive this proposal is to consider a probe brane in the action (\ref{action}). Label the embedding functions of $E$ by
\begin{eqnarray}\label{embeddingfunctionE}
X^{A}=X^{A}(y^i), 
\end{eqnarray}
the induced metric on $E$ becomes $\gamma_{ij}=\frac{\partial X^{A}}{\partial y^i}\frac{\partial X^{B}}{\partial y^j} g_{AB}(X)$. Varying the action (\ref{action}) with respect to $X^{A}$, we get
 \begin{eqnarray}\label{EOME0}
\delta_X I=T_E \int_E dy^{d+2-n} \sqrt{|\gamma|} K_{E}^{A} \ g_{AB}\delta X^{B}=0,
\end{eqnarray}
where $K_{E }^{A}=\gamma^{ij}(D_i D_j X^{A}+\Gamma^{A}_{BC} D_i X^{B} D_j X^{C})$ are the traces of extrinsic curvatures and $D_i$ denotes covariant derivatives on $E$. From (\ref{EOME0}), we read off the EOM of $X^{A}$
 \begin{eqnarray}\label{EOME}
K_{E}^{A} =0,
\end{eqnarray}
which implies that $E$ is a minimal surface in the bulk. Recall that (\ref{EOME}) is derived by $\frac{\delta}{\delta X^{A}}\int_E \sqrt{\gamma}=0$, it is clear that $E$ is a minimal surface. 

Let us go on to study the location of the end-of-world brane $Q$, which can be fixed by choosing suitable boundary conditions \cite{Takayanagi:2011zk,Miao:2018qkc}. Consider the variation of action (\ref{action}) and focus on boundary terms on $Q$, we have
 \begin{eqnarray}\label{BCforQ}
\int_Q \sqrt{|h|} \left(K^{\mu\nu}-(K-T)h^{\mu\nu}\right)\delta h_{\mu\nu}=0.
\end{eqnarray}
To have a well-defined action principle, one can impose either Neumann BC (NBC) \cite{Takayanagi:2011zk}
\begin{eqnarray}\label{NBC}
\text{NBC}:\ \ \ K^{\mu\nu}-(K-T)h^{\mu\nu}=0.
\end{eqnarray}
 or Dirichlet BC (DBC) $ \delta h_{\mu\nu}|_Q=0$ \cite{Miao:2018qkc},
 which both define a consistent theory of AdS/BCFT. Recall that, for our present case, the geometry of $Q$ are divided into two independent sectors, i.e., $\text{AdS}_{d+2-n}\times \text{S}_{n-2}$. For each sector, we can impose either NBC or DBC. For our purpose, we hope to have a dynamical gravity on the $\text{AdS}_{d+2-n}$ sector so that the cone holography can be derived from $\text{AdS}_{d+2-n}/\text{CFT}_{d+1-n}$.  This means that we should impose NBC on the $\text{AdS}_{d+2-n}$ sector \footnote{By ``$\text{AdS}_{d+2-n}$'', we means the asymptotically AdS generally. Thus the gravity can be dynamical in the $\text{AdS}_{d+2-n}$ sector. }. One choice is the NBC (\ref{NBC}) on both sectors, and other one is the mixed BC that we impose NBC on $\text{AdS}_{d+2-n}$ sector but DBC on the $ \text{S}_{n-2}$ sector. To illustrate the mixed BC, let us write the metric of Q into the following form
 \begin{eqnarray}\label{metricofQ}
ds_Q^2=h_{\mu\nu} dx^{\mu}dx^{\nu}=h_{ab} dx^a dx^b + h_{ij} dy^i dy^j,
\end{eqnarray}
where $h_{ab} $ are metrics of the $ \text{S}_{n-2}$ sector, and $h_{ij} $ are metrics for the  $\text{AdS}_{d+2-n}$ sector. Now the mixed BC (MBC) is given by
\begin{eqnarray}\label{MBC}
\text{MBC}:\ \ \ h_{ab} dx^a dx^b=l_0^2\ d\Omega_{n-2}^2, \ \ \  K^{ij}-(K-T)h^{ij}=0
\end{eqnarray}
where $l_0$ is the radius of the sphere and $d\Omega_{n-2}^2$ is the line element of $(n-2)$-dimensional unit sphere. A covariant expression of first equation of (\ref{MBC}) would be $\bar{R}_{ab}=(n-3) h_{ab}/l_0^2, \bar{C}^{abcd}=0$, where $\bar{R}_{ab}, \bar{C}^{abcd}$ denote Ricci tensor and Weyl tensor defined by $h_{ab}$ \footnote{ In addition to the local condition $\bar{R}_{ab}=(n-3) h_{ab}/l_0^2, \bar{C}^{abcd}=0$, we further require that the solution has the correct topology $\text{S}_{n-2}$.  Take the case $n=3$ as an example, we require that the solution should automatically yield a periodic angle $\phi$ rather than that one sets the period by hand.}. Below we take the MBC to illustrate the cone holography and leave the discussions of NBC to sect.4. As it will be shown below, solutions to MBC are much simpler than those to NBC. 

\subsubsection{Solutions}

Now we are ready to discuss the solutions to cone holography. It should be stressed that, as a solution to Einstein equation, the bulk geometry is smooth everywhere. The conical singularity can only appear on the defect $D$.  To warm up, let us first study the one without conical singularities on the defect $D$. We take the following ansatz of metric
\begin{eqnarray}\label{metricMBC}
ds^2=dr^2+\sinh^2( r) d\Omega_{n-2}^2+\cosh^2 (r) \bar{h}_{ij}(y)dy^i dy^j,
\end{eqnarray}
which obeys MBC (\ref{MBC}) by design and reduces to the typical metric (\ref{conemetric}) when $\bar{h}_{ij}(y)$ is an AdS metric (recall $n=m+1$). Actually $\bar{h}_{ij}(y)$ can be a more general metric.  According to \cite{Yang:2010mca}, the metric (\ref{metricMBC}) is a solution to Einstein equation with a negative cosmological constant in $d+1$ dimensions 
\begin{eqnarray}\label{EOMg}
R_{AB}-\frac{R}{2}g_{AB}=\frac{d(d-1)}{2} g_{AB},
\end{eqnarray}
provided that $\bar{h}_{ij}$ obey Einstein equation with a negative cosmological constant in $d+2-n$ dimensions
\begin{eqnarray}\label{EOMh}
R_{\bar{h}\ ij}-\frac{R_{\bar{h}}}{2}\bar{h}_{ij}=\frac{(d+1-n)(d-n)}{2} \bar{h}_{ij}.
\end{eqnarray}
Here $R_{\bar{h}\ ij}$ denote the curvatures with respect to the metric $\bar{h}_{ij}$ and we have set the AdS radius $L=1$ for simplicity.  Recall that the codim-m brane $E$, the end-of world brane $Q$ and the AdS boundary $M$ are located at $r=0$, $r=\rho$ and $r=\infty$, respectively. 

Let us explain why the metric (\ref{metricMBC}) obeys MBC  (\ref{MBC}) on the end-of-world brane $Q$ ($r=\rho$). The induced metric on $Q$ is 
 \begin{eqnarray}\label{metricofQ1}
ds_Q^2=\sinh^2( \rho) d\Omega_{n-2}^2+\cosh^2 (\rho) \bar{h}_{ij}(y)dy^i dy^j.
\end{eqnarray}
Comparing with (\ref{metricofQ}), we read off $ h_{ab} dx^a dx^b=\sinh^2( \rho)\ d\Omega_{n-2}^2$, which indeed satisfies DBC on the $S_{n-2}$ sector (\ref{MBC})  with radius $l_0=\sinh\rho$.  As for the rest part, let us calculate the extrinsic curvature $K_{\mu\nu}=\frac{1}{2} \partial_r h_{\mu\nu}|_{r=\rho}$. We get 
\begin{eqnarray}\label{KuvonQ}
K_{ab}=\coth(\rho) h_{ab},\ \ K_{ij}=\tanh(\rho) h_{ij},
\end{eqnarray}
which obeys the NBC on the $\text{AdS}_{d+2-n}$ sector as long as we parameterize the tension by $T=(n-2) \coth(\rho) +(d+1-n)\tanh(\rho)$ so that $K-T=\tanh(\rho)$ and $K_{ij}=(K-T) h_{ij}$. Now we finish the proof of the statement that the metric (\ref{metricMBC}) satisfies MBC (\ref{MBC}). 

Let us go on to explain why there is no conical singularity on the defect $D$ for the metric (\ref{metricMBC}). The arguments are similar to those around (\ref{conemetricM}). The only difference is that now $\bar{h}_{ij}$ is more general. 
For our purpose, we let $\bar{h}_{ij}$ to be the metric of an asymptotically AdS,
\begin{eqnarray}\label{metricAdShij}
\bar{h}_{ij}(y)dy^i dy^j=\frac{dz^2+\sigma_{\hat{i}\hat{j}}(z,\hat{y})dy^{\hat{i}}dy^{\hat{j}}}{z^2},
\end{eqnarray}
where $\sigma_{\hat{i}\hat{j}} =\sigma^{(0)}_{\hat{i}\hat{j}}+ z^2\sigma^{(1)}_{\hat{i}\hat{j}}+...$ From (\ref{metricMBC}) and (\ref{metricAdShij}), we notice that the induce metric on the AdS boundary $M$ ($r=\infty$) is conformally equivalent to
\begin{eqnarray}\label{conemetricM2}
ds_M^2\sim  z^2 d\Omega_{m-1}^2+dz^2+\sigma_{\hat{i}\hat{j}}(z,\hat{y})dy^{\hat{i}}dy^{\hat{j}},
\end{eqnarray}
which clearly shows that there is no conical singularity on the defect $D$ located at $z=0$.

To allow a conical singularity on the defect $D$, we choose a more general ansatz of the metric  \cite{Jensen:2013lxa,Yang:2010mca}
\begin{eqnarray}\label{metricMBCgeneral}
ds^2=dr^2+f(r) d\Omega_{n-2}^2+ g(r) \bar{h}_{ij}(y)dy^i dy^j,
\end{eqnarray}
where $ \bar{h}_{ij}$ satisfy (\ref{EOMh}).  We should impose suitable boundary conditions for $f(r)$ and $g(r)$.  As we have mentioned in sect. 2.1.2, the codim-2 brane and codim-m brane with $m\ge 3$ are quite different, we discuss them separately below. For $m=2$ ($n=3$), we choose
\begin{eqnarray}\label{conicalBC}
\lim_{r\to 0} f(r) =\frac{ r^2}{q^2},  \ \ \ \lim_{r\to 0} g(r) \ \text{is finite},  \ \ \  \lim_{r\to \infty} \frac{f(r) }{g(r)}=1,
\end{eqnarray}
where $q$ is a positive constant. Then the metric (\ref{metricMBCgeneral}) becomes 
\begin{eqnarray}\label{metricnearEm2}
ds^2\sim dr^2+\frac{r^2}{q^2} d\phi^2+...
\end{eqnarray}
near the brane $E$ ($r=0$) in the bulk, and becomes
\begin{eqnarray}\label{metricnearDm2}
ds^2\sim dz^2+z^2 d\phi^2+...
\end{eqnarray}
near the defect $D$ ($z=0$) on the AdS boundary $M$ ($r=\infty$). Here we have replaced $d\Omega_{1}^2$ by $d\phi^2$ and have used (\ref{metricAdShij}).  The period of $\phi$ is fixed to be $2\pi q$ in order to have a smooth solution (\ref{metricnearEm2})  in the bulk.  As a result, from (\ref{metricnearDm2}) there is  a conical singularity on the defect $D$ when $q\ne 1$.  Clearly, there is no way to get rid of conical singularities for both (\ref{metricnearEm2})  and (\ref{metricnearDm2}) unless $q=1$.

 As for $m>2$ ($n>3$), the situation is quite different. Remarkably, Einstein equations fix the asymptotical expression of $f(r)$ to be $\lim_{r\to 0} f(r) =r^2$, which is closely related to the fact that the tension $T_E$ must be zero for codim-m branes with $m>2$.  Please see (\ref{appansatzfg1}) and (\ref{appansatzfg2}) of the appendix for more details. Thus, we choose a different BC for this case,
 \begin{eqnarray}\label{conicalBCnne3}
\lim_{r\to 0} f(r) =r^2,  \ \ \ \lim_{r\to 0} g(r) \ \text{is finite},  \ \ \  \lim_{r\to \infty} \frac{f(r) }{g(r)}=\frac{1}{q^2}.\end{eqnarray}
Now the metric (\ref{metricMBCgeneral}) becomes 
\begin{eqnarray}\label{metricnearEm3}
ds^2\sim dr^2+r^2 d\Omega_{n-2}^2+...
\end{eqnarray}
near the brane $E$ ($r=0$) in the bulk, and becomes
\begin{eqnarray}\label{metricnearDm3}
ds^2\sim dz^2+\frac{z^2}{q^2} d\Omega_{n-2}^2+...
\end{eqnarray}
near the defect $D$ ($z=0$) on the AdS boundary $M$ ($r=\infty$).  To have a smooth bulk solution (\ref{metricnearEm3}), we choose the range of angles to be the normal ones. For example, we choose $ \theta\subset[0,\pi]$ for $d\Omega_{n-2}^2=d\theta^2+\sin^2\theta d\Omega_{n-3}^2$.  As a result, from (\ref{metricnearDm3}), there is a conical singularity on the defect $D$ for $q\ne 1$.

In general, there is no analytical solutions of (\ref{metricMBCgeneral}) when $q\ne 1$.  Fortunately,  there is an exact solution when $m=2$ ($n=3$)
\begin{eqnarray}\label{metricBHcodim3}
ds^2=\frac{d\bar{r}^2}{\bar{f}(\bar{r})}+\bar{f}(\bar{r}) d\phi^2+ \bar{r}^2 \bar{h}_{ij}(y)dy^i dy^j,
\end{eqnarray}
where $\bar{f}(\bar{r})=\bar{r}^2-1-\frac{\bar{r}_h^{d-2}}{\bar{r}^{d-2}}(\bar{r}_h^2-1)$, $\bar{r}\ge \bar{r}_h$ and 
\begin{eqnarray}\label{rbarr}
r=\int_{\bar{r}_h}^{\bar{r}} \frac{d\bar{r}}{\sqrt{\bar{f}(\bar{r})}}.
\end{eqnarray}
Imposing the conditions (\ref{conicalBC}), or equivalently, $\bar{f}'(\bar{r}_h)=2/q$, we fix the constant
\begin{eqnarray}\label{rh}
\bar{r}_h=\frac{1+\sqrt{1-2 d q^2+d^2 q^2}}{d q}.
\end{eqnarray}
Obviously, (\ref{metricBHcodim3}) is quite similar to the metric of hyperbolic black hole. The only difference is that $\phi$ is a spatial  coordinate instead of a time coordinate. The time direction is hidden in the $ \bar{h}_{ij}(y)dy^i dy^j$ sector of (\ref{metricBHcodim3}). The branes $E$ and $Q$ are located at $\bar{r}=\bar{r}_h$ and $\bar{r}=\bar{r}_0$, where $\bar{r}_0$ can be derived from MBC (\ref{MBC}) as
\begin{eqnarray}\label{Qcodim3}
K-T=\frac{\sqrt{\bar{f}(\bar{r}_0)}}{\bar{r}_0}=\tanh\rho.
\end{eqnarray}

To end this subsection, let us discuss briefly the solution to NBC (\ref{NBC}). For simplicity, let us focus on the case $m=2$ with metric (\ref{metricBHcodim3}). The codim-2 brane $E$ is still located at $\bar{r}=\bar{r}_h$ ($r=0$). However, the end-of-world brane is no longer located at constant $\bar{r}$. Instead, in order to satisfy NBC (\ref{NBC}), 
 it must depend on other coordinates, i.e., $\bar{r}=\bar{r}(\phi,y)$.   We leave a careful study of the solutions to NBC to sect.4.  Finally, it should be  mentioned that (\ref{metricMBCgeneral}) is not the most general solution to cone holography. We leave the study of more general solutions to sect.5.

\subsection{Equivalence to AdS/CFT}

In this subsection, we prove that, for the class of solutions studied in sect.2.1, the gravitational action of cone holography is equivalent  to that of AdS/CFT with Einstein gravity
\begin{eqnarray}\label{Equivalence}
I_{\text{AdSC}_{d+1}}=I_{\text{AdS}_{d+2-n}}.
\end{eqnarray}
Assuming $\text{AdS}_{d+2-n}/\text{CFT}_{d+1-n}$ holds, which means that the CFT partition function in large N limit is given by the classical gravitational action
\begin{eqnarray}\label{AdSCFT}
Z_{\text{CFT}_{d+1-n}}=e^{-I_{\text{AdS}_{d+2-n}}},
\end{eqnarray}
we get immediately a proof of cone holography
\begin{eqnarray}\label{AdSCCFT}
Z_{\text{CFT}_{d+1-n}}=e^{-I_{\text{AdSC}_{d+1}}},
\end{eqnarray}
at least for the class of solutions of sect.2.1. Note that $I_{\text{AdS}_{d+2-n}}$ and $I_{\text{AdSC}_{d+1}}$ of (\ref{AdSCFT},\ref{AdSCCFT}) are Euclidean actions. For simplicity, below we focus on the actions in Lorentz signature, which differ from the ones in Euclidean signature by a minus sign. 

To warm up, let us first consider the case without conical singularities, i.e., $q=1$.  Equivalently, the tension of the codim-m brane is zero, i.e., $T_E=0$. Recall that $T_E$ always vanishes for codim-m branes with $m>2$ and $T_E\sim (q-1)/q=0$ for codim-2 branes when $q=1$.  On the contrary, the tension of codim-1 brane $Q$ is non-zero, i.e., $T\ne 0$. 
Substituting the metric (\ref{metricMBC}) into the action (\ref{action}) and applying MBC (\ref{MBC}) together with the formula (\ref{curvature7}) with $f(r)=\sinh^2(r), g(r)=\cosh^2(r)$, we derive
\begin{eqnarray}\label{AdSCaction1}
I_{\text{AdSC}_{d+1}}
&=&\frac{V_{S_{n-2}}}{16\pi G_N}\int_0^{\rho} \sinh^{n-2} (r)\cosh^{d+2-n}(r)dr \nonumber\\
&&\times \int_{\bar{Q}} \sqrt{|\bar{h}|} \Big{(} R_{\bar{h}} \text{sech}^2(r)+(d-n+1) (d-n+2) \text{sech}^2(r)-2d \Big{)}\nonumber\\
&&+\frac{V_{S_{n-2}}}{8\pi G_N}\int_{\bar{Q}} \sqrt{|\bar{h}|} \sinh^{n-2} (\rho)\cosh^{d+2-n}(\rho) \tanh\rho \nonumber\\
&=& \frac{V_{S_{n-2}}}{16\pi G_N}\int_0^{\rho} \sinh^{n-2} (r)\cosh^{d-n}(r) dr \int_{\bar{Q}} \sqrt{|\bar{h}|} \Big{(} R_{\bar{h}} +(d+1-n)(d-n) \Big{)} \nonumber\\
&=& \frac{1}{16\pi G^{(d+2-n)}_N}\int_{\bar{Q}} \sqrt{|\bar{h}|} \Big{(} R_{\bar{h}} +(d+1-n)(d-n) \Big{)}=I_{\text{AdS}_{d+2-n}}\ ,
\end{eqnarray}
which is equal to the gravitational action $I_{\text{AdS}_{d+2-n}}$ with Newton's constant given by
\begin{eqnarray}\label{Newton's constant1}
\frac{1}{G^{(d+2-n)}_N}=\frac{V_{S_{n-2}}}{G_N}\int_0^{\rho} \sinh^{n-2} (r)\cosh^{d-n}(r) dr,
\end{eqnarray}
where $\bar{Q}$ denotes the $\text{AdS}_{d+2-n}$ sector of $Q$ and $V_{S_{n-2}}=\frac{2 \pi ^{\frac{n-1}{2}}}{\Gamma \left(\frac{n-1}{2}\right)}$ is the volume of (n-2)-dimensional unit sphere. 
Note that we take $\bar{h}_{ij}$ off-shell in the above derivations, which means that $\bar{h}_{ij}$ need not satisfy (\ref{EOMh}). Besides, we have used $-2\Lambda=d(d-1)$, $K-T=\tanh\rho$ and the following formula
\begin{eqnarray}\label{integralformula}
\int_0^{\rho} \sinh ^{n-2}(r) \cosh ^{d-n}(r) \left(d-2 n+2-d\cosh (2 r)\right) dr=- 2 \sinh ^{n-1}(\rho ) \cosh ^{d-n+1}(\rho ).
\end{eqnarray}
Now we have proved the equivalence (\ref{Equivalence}) between cone holography and AdS/CFT for the case without conical singularities.

Let us go on to discuss the case with non-trivial conical singularities, i.e., $q\ne 1$. For simplicity, let us first consider the codim-3 holography, which has an exact solution (\ref{metricBHcodim3}). Note that the solution (\ref{metricBHcodim3}) is smooth everywhere in the bulk, even at the location of $E$, the real conical singularity is on the conical defect $D$ instead of $E$. As a result, one does not count the action of codim-2 brane $I_E=-T_E\int_E \sqrt{\gamma}$ when calculating the gravitational action in the bulk \cite{Dong:2016fnf,Hung:2011nu,Dong:2016wcf}. It is more like a trick to determine the location of $E$ but does not contribute to the action directly \footnote{ It affects the action by the backreaction to the bulk solution. }. Another way to understand this is that we regularize the integral region by $r\ge \epsilon>0$ and taking the limit $\epsilon \to 0$ at the end of calculations. These are the common methods used to study holographic R\'enyi entropy, where there is a cosmic brane in the bulk \cite{Dong:2016fnf,Hung:2011nu,Dong:2016wcf}.  Another example is that, to calculate the holographic free energy, one does not take into account the contribution of ``conical singularity'' on the horizon of black hole. The spacetime is smooth, and nothing special happens on the horizon. 

Substituting (\ref{metricBHcodim3}) into the action (\ref{action}) without $I_E$ and applying boundary condition (\ref{Qcodim3}) and  $\bar{f}(\bar{r})=\bar{r}^2-1-\frac{\bar{r}_h^{d-2}}{\bar{r}^{d-2}}(\bar{r}_h^2-1)$, we obtain
\begin{eqnarray}\label{AdSCaction2}
I_{\text{AdSC}_{d+1}}
&=&\frac{2\pi q}{16\pi G_N}\int_{\bar{r}_h}^{\bar{r}_0} d\bar{r} \bar{r}^{d-1} \sqrt{|\bar{h}|}\left(\frac{R_{\bar{h}}+(d-1) (d-2)}{\bar{r}^2}-2 d\right)\nonumber\\
&&+\frac{2\pi q}{8\pi G_N}\int_{\bar{Q}} \sqrt{|\bar{h}|} \bar{r}_0^{d-2} \bar{f}(\bar{r}_0) \nonumber\\
&=& \frac{q}{8 G_N} \frac{\bar{r}_0^{d-2}-\bar{r}_h^{d-2}}{d-2} \int_{\bar{Q}} \sqrt{|\bar{h}|} \Big{(} R_{\bar{h}} +(d-2)(d-3) \Big{)} \nonumber\\
&=& \frac{1}{16\pi G^{(d-1)}_N}\int_{\bar{Q}} \sqrt{|\bar{h}|} \Big{(} R_{\bar{h}} +(d-2)(d-3) \Big{)}=I_{\text{AdS}_{d-1}}\ ,
\end{eqnarray}
which proves the equivalence (\ref{Equivalence}) with $n=3$ ($m=2$) for general $q$, provided that the effective Newton's constant on $\bar{Q}$ is defined by
\begin{eqnarray}\label{Newton's constant2}
\frac{1}{G^{(d-1)}_N}=\frac{2\pi q}{G_N} \frac{(\bar{r}_0^{d-2}-\bar{r}_h^{d-2})}{d-2},
\end{eqnarray}
where recall that $q=2/\bar{f}'(\bar{r}_h)=2 \bar{r}_h/(d \bar{r}_h^2-d+2)$.

Now let us consider the most general case, the cone holography with general codimensions $n=m+1$ and non-trivial conical singularities, i.e., $q\ne 1$.  We focus on the class of solution (\ref{metricMBCgeneral}) with $0 \le r \le r_0$.  Following the above approach, we derive
\begin{eqnarray}\label{AdSCaction3}
I_{\text{AdSC}_{d+1}}
&=&\frac{V_{\hat{S}_{n-2}}}{16\pi G_N}\int_{0}^{r_0} dr f(r)^{\frac{n-2}{2}}  g(r)^{\frac{d+2-n}{2}}  \sqrt{|\bar{h}|}\left(\frac{R_{\bar{h}}+(d+2-n) (d+1-n)}{g(r)}-2 d\right)\nonumber\\
&&+\frac{V_{\hat{S}_{n-2}}}{16\pi G_N}\int_{\bar{Q}} \sqrt{|\bar{h}|} f(r_0)^{\frac{n-2}{2}}  g(r_0)^{\frac{d-n}{2}}  g'(r_0) \nonumber\\
&=& \frac{1}{16\pi G^{(d-1)}_N}\int_{\bar{Q}} \sqrt{|\bar{h}|} \Big{(} R_{\bar{h}} +(d+1-n)(d-n) \Big{)}=I_{\text{AdS}_{d+2-n}}\ ,
\end{eqnarray}
where the Newton's constants are related by
\begin{eqnarray}\label{Newton's constant3}
\frac{1}{G^{(d+2-n)}_N}=
\frac{V_{\hat{S}_{n-2}}}{G_N}\int_{0}^{r_0} dr f(r)^{\frac{n-2}{2}}  g(r)^{\frac{d-n}{2}},
\end{eqnarray}
and $r=r_0$ is the location of end-of world brane $Q$. Note that $V_{\hat{S}_{n-2}}=2\pi q$ for $n=3$ and $V_{\hat{S}_{n-2}}=V_{S_{n-2}}$ for $n>3$.
To derive (\ref{AdSCaction3}), we have used the integral formula 
\begin{eqnarray}\label{integralformulageneral}
\int_0^{r_0} f(r)^{\frac{n-2}{2}}g(r) ^{\frac{d-n}{2}}\left(n-1-d+d g(r)\right) dr=\frac{1}{2} f(r_0)^{\frac{n-2}{2}}g(r_0)^{\frac{d-n}{2}}g'(r_0).
\end{eqnarray}
It is interesting that, although the exact expressions of $f(r)$ and $g(r)$ are un-known generally, the EOM and BC of  $f(r)$ and $g(r)$ are sufficient to derive (\ref{integralformulageneral}). The proof is as follows. Differentiating (\ref{integralformulageneral}) with respect to $r_0$, we get
\begin{eqnarray}\label{proofintegralgeneral1}
4 d g(r)^2-2 g(r) \left(2 d+g''(r)-2 n+2\right)+(n-d) g'(r)^2-\frac{(n-2) g(r) f'(r) g'(r)}{f(r)}=0,
\end{eqnarray}
which is just the EOM (\ref{EOMfg3}). Note that we have replace $r_0$ by $r$ above.  To prove (\ref{integralformulageneral}), we still need to verify that 
\begin{eqnarray}\label{proofintegralgeneral2}
f(0)^{\frac{n-2}{2}}g(0)^{\frac{d-n}{2}}g'(0)=0.
\end{eqnarray}
Recall the BC (\ref{conicalBC},\ref{conicalBCnne3}) of $f(r)$ and $g(r)$, which yields $\lim_{r\to 0}f(r)^{\frac{n-2}{2}}g(r)^{\frac{d-n}{2}}g'(r)\sim O(r^{n-2})=0$. So (\ref{proofintegralgeneral2}) is indeed satisfied.  Now we finish the proof of (\ref{integralformulageneral}) by using EOM and BC of $f(r)$ and $g(r)$.

In the above discussions, we focus on the non-renormalized actions (\ref{AdSCaction1},\ref{AdSCaction2},\ref{AdSCaction3}), which are divergent generally. To get finite results, one can perform the holographic renormalization \cite{Balasubramanian:1999re,deHaro:2000vlm} by adding suitable counterterms on the defect $D$.  Following \cite{Balasubramanian:1999re,deHaro:2000vlm}, we choose the following counterterms on $D$
\begin{eqnarray}\label{counterterms}
I_{\text{counter}}=\frac{1}{16\pi G^{(d)}_N} \int_{D} \sqrt{|\sigma|} \left( 2 K_{\Sigma} +2(n-d)-\frac{1}{d-1-n} R_{\Sigma}+...\right),
\end{eqnarray}
which makes the equivalence
\begin{eqnarray}\label{keyequivalence}
I_{\text{AdSC}_{d+1}}+I_{\text{counter}}=I_{\text{AdS}_{d+2-n}}+I_{\text{counter}}
\end{eqnarray}
still holds after renormalization. 
Now we finish the proof of the statement that $\text{AdSC}_{d+1}/\text{CFT}_{d+1-n}$ with the solutions (\ref{metricMBC},\ref{metricMBCgeneral},\ref{metricBHcodim3}) is equivalent to $\text{AdS}_{d+2-n}/\text{CFT}_{d+1-n}$ with Einstein gravity, at least at the classical level for gravity, or equivalently, in the large N limit for CFTs. 

 The equivalence (\ref{keyequivalence}) is quite powerful which enables us to derive many interesting physical quantities such as Entanglement/R\'enyi entropy for cone holography directly following the approach of AdS/CFT. See sect.3 for examples. Assuming that AdS/CFT holds which is widely accepted, the equivalence (\ref{keyequivalence}) is actually a proof of the cone holography in a certain sense. It should be stressed that the solution (\ref{metricBHcodim3}) is not the most general solution to cone holography. As a result, in general, cone holography is different from AdS/CFT with Einstein gravity. That is because there are infinite towers of massive Kaluza-Klein modes on the branes. And the effective gravity on the brane is massive gravity instead of Einstein gravity generally. This is consistent with the interpretation of cone holography as a holographic dual of edge modes. As mentioned in the introduction, the edge modes include bulk information and differ from the usual CFTs. Thus, in general, the cone holography as a holographic dual of edge modes is different from AdS/CFT.

\section{Aspects of cone holography}

Since cone holography with the solutions (\ref{metricMBCgeneral}) is equivalent to AdS/CFT with vacuum Einstein gravity, many interesting results of AdS/CFT can be reproduced in cone holography. These include holographic Weyl anomaly, holographic Entanglement/R\'enyi entropy and holographic correlation functions, which all agree with the results of CFTs.  See \cite{Miao:2020oey} for example, where the case of wedge holography is carefully studied.  The generalization to cone holography is straightforward. Actually, we only need to replace the Newton's constant $G^{d}_N$ of \cite{Miao:2020oey}  by  $G^{d+2-n}_N$ (\ref{Newton's constant3}) for cone holography with the class of solutions (\ref{metricMBCgeneral}).  Thus, we do not repeat the calculations here. Instead, we only list some of the key results and steps for the convenience of readers.  For simplicity, let us focus on the cone holography $\text{AdSC}_{n+2}/\text{CFT}_{2}$ below.

\subsection{Holographic Weyl anomaly}

We assume that the spacetime on $E$ and $\bar{Q}$ is an asymptotically AdS
\begin{eqnarray}\label{metricAdS3}
ds^2=dr^2+\sinh^2( r) d\Omega_{n-2}^2 +\cosh^2 (r) \frac{dz^2+\sigma_{\hat{i}\hat{j}}dy^{\hat{i}}dy^{\hat{j}}}{z^2},
\end{eqnarray}
where $\sigma_{\hat{i}\hat{j}} =\sigma^{(0)}_{\hat{i}\hat{j}}+ z^2(\sigma^{(1)}_{\hat{i}\hat{j}} +\lambda^{(1)}_{\hat{i}\hat{j}}\ln z)+...$, and $\sigma^{(0)}_{\hat{i}\hat{j}}$ is the metric on defect $D$. Solving Einstein equations (\ref{EOMg}) , we get
\begin{eqnarray}\label{g12d}
\sigma^{(0)\hat{i}\hat{j}}\sigma^{(1)}_{\hat{i}\hat{j}}=-\frac{R_{D}}{2},
\end{eqnarray}
where $R_D$ is Ricci scalar on the defect $D$. Note that (\ref{g12d}) can also be obtained  from the asymptotical symmetry of AdS \cite{Imbimbo:1999bj}, which plays an important role in the off-shell derivations of holographic Weyl anomaly \cite{Miao:2013nfa}. Substituting the above two equations into the gravitational action (\ref{action}) and select the UV logarithmic divergent term, we can derive the holographic Weyl anomaly \cite{Henningson:1998gx}. We get
\begin{eqnarray}\label{Weylanomaly2d}
\mathcal{A}=\int_{D} dy^{2}\sqrt{|\sigma|} \frac{c}{24 \pi} R_{D},
\end{eqnarray}
with the central charge 
\begin{eqnarray}\label{charge2d}
c=\frac{3V_{S_{n-2}}}{2G_N}\int_0^{\rho} \sinh^{n-2} (r)\cosh(r) dr=\frac{3\pi ^{\frac{n-1}{2}}}{G_N\Gamma \left(\frac{n-1}{2}\right)}\frac{\sinh ^{n-1}(\rho)}{n-1}.
\end{eqnarray}
It is interesting that the central charge $c$ is a monotonically increasing function of $\rho$
\begin{eqnarray}\label{ctheorem0}
\partial_{\rho} c\ge 0.
\end{eqnarray}
Recall that $r=\rho$ denotes the location of end-of-world brane $Q$. The larger $\rho$ is, the closer the brane $Q$ tends into AdS boundary $M$. See Fig.\ref{Figureconemetric} for example. Note that the AdS boundary corresponds to UV, while the deep bulk corresponds to IR. Thus, we have 
\begin{eqnarray}\label{ctheorem1}
\rho_{UV} > \rho_{IR}.
\end{eqnarray}
It should be mentioned that (\ref{ctheorem1}) can also be derived from the null energy condition on $Q$ \cite{Fujita:2011fp}. From (\ref{ctheorem0}) and (\ref{ctheorem1}), we get a holographic proof of the c-theorem \cite{Cardy:1988cwa,Zamolodchikov:1986gt}
\begin{eqnarray}\label{ctheorem2}
c_{UV} \ge  c_{IR}.
\end{eqnarray}
This is a strong support for our proposal of cone holography. 

The above discussions apply to the case without conical singularities, i.e., $q= 1$. It is straightforward to extend the above discussions to general case with $q \ne 1$. By applying the solution (\ref{metricMBCgeneral}), we obtain the Weyl anomaly (\ref{Weylanomaly2d}) with the central charge 
\begin{eqnarray}\label{charge2dgeneral}
c=\frac{3V_{\hat{S}_{n-2}}}{2G_N}\int_0^{r_0} f(r)^{\frac{n-2}{2}} g(r)^{\frac{1}{2}} dr,
\end{eqnarray}
where $r_0$ is the location of $Q$. Since $f(r)$ and $g(r)$ are positive functions, we have $\partial_{r_0} c\ge 0$. Following the above arguments, we have $r_{0 UV}\ge r_{0 IR}$. Thus the c-theorem (\ref{ctheorem2}) is still obeyed for the general case with $q \ne 1$. 

\subsection{Holographic R\'enyi entropy}

R\'enyi entropy measures the quantum entanglement of a subsystem, which is defined by 
\begin{eqnarray}\label{Renyientropy}
S_{p}=\frac{1}{1-p} \ln \text{tr} \rho_A^{p},
\end{eqnarray}
where $p$ is a positive integer, $\rho_A=\text{tr}_{\bar{A}}\  \rho$ is the induced density matrix of a subregion $A$. Here $\bar{A}$ denotes the complement of $A$ and $\rho$ is the density matrix of the whole system.  In the limit $p\to 1$, R\'enyi  entropy becomes the von Neumann entropy, which is also called entanglement entropy
\begin{eqnarray}\label{Entanglemententropy}
S_{\text{EE}}=-\text{tr} \rho_A \ln \rho_A.
\end{eqnarray}
In the gravity dual, R\'enyi entropy can be calculated by the area of a codim-2 cosmic brane \cite{Dong:2016fnf}
\begin{eqnarray}\label{HoloRenyi}
{p}^2\partial_{p}\left( \frac{p-1}{p} S_{p} \right)=\frac{\text{Area(Cosmic Brane}_{p})}{4 G_N},
\end{eqnarray}
where the cosmic brane${}_p$ is anchored at the entangling surface $\partial A$.  Since the tension of cosmic brane
$T_{p}=\frac{p-1}{4p G_N}$ is non-zero generally, it backreacts on the bulk geometry.  In the tensionless limit $p\to 1$, the cosmic brane becomes a minimal surface and (\ref{HoloRenyi}) becomes the Ryu–Takayanagi formula for entanglement entropy \cite{Ryu:2006bv}
\begin{eqnarray}\label{HoloEE}
S_{\text{EE}}=\frac{\text{Area(Minimal Surface)}}{4 G_N}.
\end{eqnarray}

For cone holography, the holographic R\'enyi entropy is still given by (\ref{HoloRenyi}). What is new is that the codim-2 cosmic brane ends on the end-of-world brane $Q$ and codim-m brane $E$. 
 The location of cosmic brane can be fixed by solving Einstein equations with backreactions  \cite{Dong:2016fnf}.  

Inspired by \cite{Hung:2011nu,Dong:2016wcf,Chu:2016tps}, we make the following ansatz of the bulk metric 
 \begin{eqnarray}\label{metricRenyirh}
ds^2=dr^2+\sinh^2( r) d\Omega_{n-2}^2 +\cosh^2 (r) \left( \frac{d\tilde{r}^2}{\tilde{r}^2-\frac{1}{p^2}} -(\tilde{r}^2-\frac{1}{p^2})dt^2+\tilde{r}^2 d H_{1}^2  \right)
\end{eqnarray}
where $p$ is the R\'enyi index, $d H_{1}^2=dy^2/y^2$ is the line element of one-dimensional hyperbolic space. The cosmic brane is just the horizon of hyperbolic black hole, whose area is given by
\begin{eqnarray}\label{Renyiarea}
\text{Area(Cosmic Brane}_{p})=\frac{V_{S_{n-2}}V_{H_{1}}}{p} \int_0^{\rho}\sinh^{n-2}(r) \cosh(r)dr,
\end{eqnarray}
where $ V_{H_1}$ is the volume of hyperbolic space.  From (\ref{HoloRenyi},\ref{metricRenyirh},\ref{Renyiarea}), we finally obtain the holographic R\'enyi entropy for $\text{CFT}_2$ as
\begin{eqnarray}\label{Renyifinalresult}
S_{p}=\frac{p+1}{ p} \frac{V_{S_{n-2}} V_{H_{1}}}{8 G_N}\int_0^{\rho}\sinh^{n-2}(r) \cosh(r)dr=\frac{p+1}{ p} \frac{c}{12}V_{H_{1}},
\end{eqnarray}
where we have used (\ref{charge2d}) above. 
Note that $V_{H_{1}}$ includes a log term $V_{H_1}|_{\ln \frac{1}{\epsilon}} =2$ \cite{Hung:2011nu},
(\ref{Renyifinalresult}) gives the correct universal term of Renyi entropy 
\begin{eqnarray}\label{Renyifinalresult1}
S_{p}|_{\ln \frac{1}{\epsilon}}=\frac{p+1}{ p} \frac{c}{6}
\end{eqnarray}
This is also a support for the cone holography. 

\subsection{Holographic correlation functions}

In this subsection, we study the correlation functions for cone holography. The holographic two point functions of stress tensors can be derived following the approach of \cite{Miao:2020oey} for wedge holography.  We do not repeat it here. Instead, we study the two point functions of scalar operators for cone holography $\text{AdSC}_{n+2}/\text{CFT}_{2}$.

Let us focus on the probe limit with the bulk metric given by
\begin{eqnarray}\label{probelimit}
ds^2=\frac{d\bar{r}^2}{\bar{r}^2-1}+(\bar{r}^2-1)d\Omega_{n-2}^2+\bar{r}^2 ds^2_{\text{AdS}_3},
\end{eqnarray}
which is a locally AdS space. 
We make the following ansatz of bulk scalar
\begin{eqnarray}\label{bulkscalar}
\phi=\phi_{\bar{r}}(\bar{r}) \phi_y(y)
\end{eqnarray}
where $y$ denotes the coordinate of $ds^2_{\text{AdS}_3}$.  Substituting (\ref{bulkscalar}) into the Klein-Gordon equation in the bulk
\begin{eqnarray}\label{KGequation}
\frac{\partial_A(\sqrt{g}g^{AB}\partial_B \phi)}{\sqrt{g}}-\hat{M}^2 \phi=0,
\end{eqnarray}
we get 
\begin{eqnarray}\label{EOMscalar}
\frac{\partial_{\bar{r}}\left( \bar{r}^3 (\bar{r}^2-1)^{\frac{n-1}{2}}\partial_{\bar{r}} \phi_{\bar{r}}(\bar{r}) \right)}{\bar{r}^3 (\bar{r}^2-1)^{\frac{n-3}{2}}\phi_{\bar{r}}(\bar{r}) }+\frac{\Box_{y}  \phi_y(y)}{\bar{r}^2\phi_y(y)}=\hat{M}^2,
\end{eqnarray}
where $\Box_{y}$ denotes D 'Alembert operator in $\text{AdS}_3$. 
To solve (\ref{EOMscalar}), we assume $\phi_y(y)$ satisfies Klein-Gordon equation in $\text{AdS}_3$
\begin{eqnarray}\label{KGequationAdS3}
\Box_{y}  \phi_y(y)- \hat{m}^2  \phi_y(y)=0,
\end{eqnarray}
where $\hat{m}$ is a constant and will be determined later. 
Then (\ref{EOMscalar}) becomes
\begin{eqnarray}\label{EOMscalar1}
\frac{\partial_{\bar{r}}\left( \bar{r}^3 (\bar{r}^2-1)^{\frac{n-1}{2}}\partial_{\bar{r}} \phi_{\bar{r}}(\bar{r}) \right)}{\bar{r}^3 (\bar{r}^2-1)^{\frac{n-3}{2}} }=(\hat{M}^2-\frac{\hat{m}^2}{\bar{r}^2})\phi_{\bar{r}}(\bar{r}),
\end{eqnarray}
which can be solved as
\begin{eqnarray}\label{solutionscalar}
\phi_{\bar{r}}&=&c_1 r^{-\hat{m}} \, _2F_1\left[\frac{ -2 \hat{m}+n-\sqrt{4 \hat{M}^2+(n-1)^2}-1}{4},\frac{ -2 \hat{m}+n+\sqrt{4 \hat{M}^2+(n-1)^2}-1}{4};1-\hat{m};r^2\right]\nonumber\\
&+&c_2 r^{\hat{m}} \, _2F_1\left[\frac{2 \hat{m}+n-\sqrt{4 \hat{M}^2+(n-1)^2}-1}{4} ,\frac{2 \hat{m}+n+\sqrt{4 \hat{M}^2+(n-1)^2}-1}{4} ;\hat{m}+1;r^2\right],\nonumber\\
\end{eqnarray}
where $\, _2F_1$ is the hypergeometric function, $c_1$ and $c_2$ are integral constants.  Without loss of generality, we can set
\begin{eqnarray}\label{c1scalar}
c_1=1.
\end{eqnarray}
We impose the natural boundary condition on the codim-(n-1) brane $E$,
\begin{eqnarray}\label{BConE}
\lim_{\bar{r}\to 1} \phi_{\bar{r}}(\bar{r}) \ \text{is finite},
\end{eqnarray}
which yields
\begin{eqnarray}\label{c2scalarodd}
c_2=-\frac{\Gamma (1-\hat{m}) \Gamma \left(\frac{2 \hat{m}+5-n-\sqrt{4 \hat{M}^2+(n-1)^2}}{4} \right) \Gamma \left(\frac{2 \hat{m}+5-n+\sqrt{4 \hat{M}^2+(n-1)^2}}{4}\right)}{\Gamma (\hat{m}+1) \Gamma \left(\frac{-2 \hat{m}+5-n-\sqrt{4 \hat{M}^2+(n-1)^2}}{4} \right) \Gamma \left(\frac{-2 \hat{m}+5-n+\sqrt{4 \hat{M}^2+(n-1)^2}}{4}\right)}
\end{eqnarray}
for odd $n$, and 
\begin{eqnarray}\label{c2scalareven}
c_2=-\frac{\Gamma (1-\hat{m}) \Gamma \left(\frac{2 \hat{m}+n-1-\sqrt{4 \hat{M}^2+(n-1)^2}}{4} \right) \Gamma \left(\frac{2 \hat{m}+n-1+\sqrt{4 \hat{M}^2+(n-1)^2}}{4}\right)}{\Gamma (\hat{m}+1) \Gamma \left(\frac{-2 \hat{m}+n-1-\sqrt{4 \hat{M}^2+(n-1)^2}}{4} \right) \Gamma \left(\frac{-2 \hat{m}+n-1+\sqrt{4 \hat{M}^2+(n-1)^2}}{4}\right)}
\end{eqnarray}
for even $n$. 
On the end-of-world brane $Q$, we can impose either DBC 
\begin{eqnarray}\label{DBCscalar}
 \phi_{\bar{r}}(\bar{r}_0)=0,
\end{eqnarray}
or NBC
\begin{eqnarray}\label{NBCscalar}
\partial_{\bar{r}} \phi_{\bar{r}}(\bar{r}_0)=0,
\end{eqnarray}
where $\bar{r}_0$ is the location of $Q$.  For simplicity, we do not show the exact expressions of (\ref{DBCscalar}) and (\ref{NBCscalar}). From (\ref{DBCscalar}) or (\ref{NBCscalar}), in principle, we can solve $\hat{m}$ in terms of $\hat{M}$
\begin{eqnarray}\label{mMn}
\hat{m}=\hat{m}(\hat{M},n,\bar{r}_0).
\end{eqnarray}
There are infinite solutions for the allowed mass $\hat{m}$, which corresponds to the infinite massive KK modes. 

Now we have fixed the mass $\hat{m}$ from boundary conditions and we are ready to derive the two point function of scalar operators. According to AdS/CFT, the bulk scalar field with mass $\hat{m}$ in $\text{AdS}_3$ is dual to a scalar operator $O$ with the conformal dimension
\begin{eqnarray}\label{conformaldimension}
\Delta=1+\sqrt{1+\hat{m}^2},
\end{eqnarray}
and the two point functions of $O$ is given by
\begin{eqnarray}\label{twopointscalar}
<O(x)O(x')>=\frac{1}{|x-x'|^{2\Delta}}.
\end{eqnarray}

Let us show more details for the derivation of holographic two point function (\ref{twopointscalar}) for cone holography.  The  bulk scalar action is given by
\begin{eqnarray}\label{actionscalar}
I=-\frac{1}{2} \int_C \sqrt{|g|} \left(  g^{AB} \partial_A \phi \partial_B \phi + \hat{M}^2 \phi^2  \right).
\end{eqnarray}
Substituting the ansatz (\ref{bulkscalar}) into (\ref{actionscalar}) and using EOM (\ref{EOMscalar1}) together with BCs (\ref{DBCscalar},\ref{NBCscalar}), we can obtain
\begin{eqnarray}\label{actionscalar1}
I&=&-\frac{1}{2} \int_C \sqrt{|g|} \left(  g^{{\bar{r}} {\bar{r}} } (\partial_{\bar{r}} \phi_{\bar{r}})^2   \phi_y^2+g^{ij}\partial_i\phi_y\partial_j\phi_y \phi_{\bar{r}}^2 + \hat{M}^2 \phi_{\bar{r}}^2\phi_y^2  \right)\nonumber\\
&=& -\frac{1}{2} \int_C \sqrt{|g|} \left(\frac{-1}{ \sqrt{|g|} }\partial_{\bar{r}} (\sqrt{|g|}  g^{{\bar{r}} {\bar{r}} } \partial_{\bar{r}} \phi_{\bar{r}})  \phi_{\bar{r}}  \phi_y^2+g^{ij}\partial_i\phi_y\partial_j\phi_y \phi_{\bar{r}}^2 + \hat{M}^2 \phi_{\bar{r}}^2\phi_y^2  \right)\nonumber\\
&=&-\frac{V_{S_{n-2}}}{2}\int_{1}^{\bar{r}_0} \bar{r}(\bar{r}^2-1)^{\frac{n-3}{2}} \phi_{\bar{r}}^2(\bar{r})\ d\bar{r} \int \sqrt{|\bar{h}|} (\bar{h}^{ij} \partial_i\phi_y\partial_j\phi_y+ \hat{m}^2 \phi_y^2) \nonumber\\
&\sim &  -\frac{1}{2} \int dy^3 \sqrt{|\bar{h}|} (\bar{h}^{ij} \partial_i\phi_y\partial_j\phi_y+ \hat{m}^2 \phi_y^2),
\end{eqnarray}
which is proportional to the scalar action in $\text{AdS}_3$ up to a constant factor. Recall that $\bar{h}_{ij}$ is the metric of $\text{AdS}_3$. Now following the standard approach of AdS/CFT \cite{Witten:1998qj}, we can derive the two point function (\ref{twopointscalar}) from (\ref{actionscalar1}).

To summary, we have shown that the cone holography can produce the correct Weyl anomaly, Entanglement/ R\'enyi entropy and correlation functions, which are strong supports for our proposal. For simplicity, we focus on  $\text{AdSC}_{n+2}/\text{CFT}_{2}$ in this section. Following \cite{Miao:2020oey}, the generalization to $\text{AdSC}_{d+1}/\text{CFT}_{d+1-n}$ is straightforward.

\section{Cone holography with Neumann BC}

In this section, we discuss the cone holography with NBC. Compared with MBC and DBC, it is more difficult to find solutions to NBC generally. For example, as mentioned at the end of sect.2.1, the natural embedding function $r=r_0$ of brane $Q$ does not satisfy NBC \footnote{ It obeys NBC only if $r=r_0 \to \infty$, which means that the end-of-world brane $Q$ approaches the AdS boundary $M$.} . To satisfy NBC, we can consider more general embedding functions. For simplicity, we focus on $\text{AdSC}_5/\text{CFT}_2$ below, which is the simplest non-trivial example.

The bulk metric for $\text{AdSC}_5/\text{CFT}_2$  is given by (\ref{metricBHcodim3}) with $d=4$.  The codim-2 brane $E$ is at $\bar{r}=\bar{r}_h$ (\ref{rh}), and the embedding function of the codim-1 brane $Q$ is assumed to be
\begin{eqnarray}\label{embeddingQNBC}
\bar{r}=F(\phi).
\end{eqnarray}
Imposing NBC (\ref{NBC}), we derive one independent equation
\begin{eqnarray}\label{NBCsect41}
F'(\phi)=\pm \frac{\bar{f}(F(\phi )) \sqrt{\coth ^2\rho \ \bar{f}(F(\phi ))-F(\phi )^2}}{F(\phi )},
\end{eqnarray}
where recall that $\bar{f}(\bar{r})=\bar{r}^2-1-\frac{\bar{r}_h^2}{\bar{r}^2}(\bar{r}_h^2-1)$.  Without loss of generality, we focus on the case $F'(\phi)\ge 0$ below. The other case with  $F'(\phi)\le 0$ can be obtained from the one with  $F'(\phi)\ge 0$ by the symmetry $\phi\to -\phi$. 

\begin{figure}[t]
\centering
\includegraphics[width=12cm]{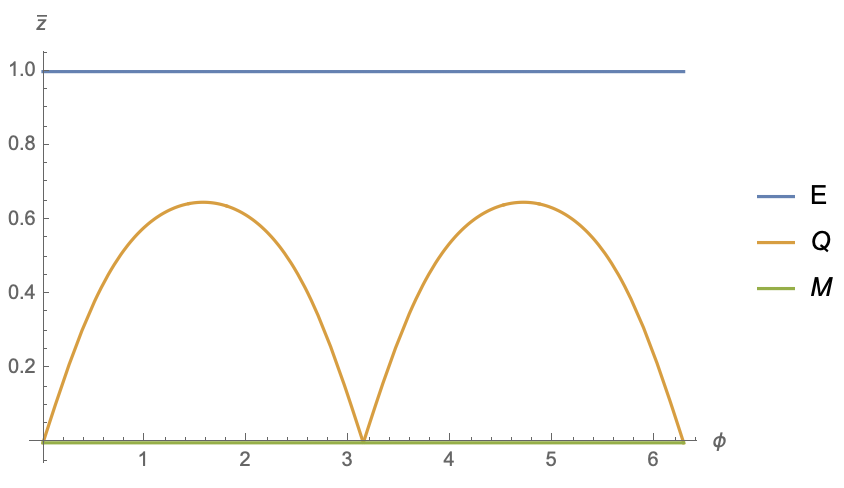}
\caption{ Cone holography with NBC for $\bar{r}_h=1$ ($q=1$). The codim-2 brane $E$ ($\bar{z}=1$), codim-1 brane $Q$ (\ref{Qcodim3NBCm1}) and AdS boundary $M$  ($\bar{z}=0$) are labelled by the blue, orange and green lines, respectively.  The bulk metric is given by  (\ref{metricBHcodim3}) with $d=4$ and $\bar{r}_h=1$. The embedding function of $Q$ is given by (\ref{Qcodim3NBCm1}). Note that $\bar{z}=1/\bar{r}$ and $\phi \simeq \phi+2\pi$. Note also that the defect $D$ with $z=0$ (\ref{metricAdShij}) is not shown in this figure. }
\label{Figureconeangle}
\end{figure}

For $\bar{r}_h=1$,  we can solve
 \begin{eqnarray}\label{Qcodim3NBCm1}
\bar{r}=F(\phi)=\sqrt{\cosh^2\rho+\sinh^2\rho \ \cot^2\phi}.
\end{eqnarray}
 Note that $\bar{r}=\infty$ for $\phi=0, \pi$, which means that $Q$  intersects the AdS boundary $M$ at these two angles.  See Fig.\ref{Figureconeangle} for example, where $\bar{z}=1/\bar{r}$.

For general $\bar{r}_h$, there is no analytical solution to (\ref{NBCsect41}). Instead, we get a integral expression
\begin{eqnarray}\label{NBCthetasolution}
\phi(\bar{r})=\int_{F_{\text{min}}}^{\bar{r}} \frac{F dF}{\bar{f}(F) \sqrt{\coth ^2\rho \ \bar{f}(F)-F^2}} ,
\end{eqnarray}
where
\begin{eqnarray}\label{Fmin}
F_{\text{min}}=\frac{1}{2} \sqrt{2 \cosh ^2(\rho )+\sqrt{2} \cosh (\rho ) \sqrt{\cosh (2 \rho )+8 \bar{r}_h^4-8 \bar{r}_h^2+1}},
\end{eqnarray}
which is derived from $\sqrt{\coth ^2\rho \ f(F_{\text{min}})-F_{\text{min}}^2}=0$.  Note that (\ref{NBCthetasolution}) can not cover the full range of the angle, i.e., $2\pi q $. Instead, it only covers the piece
 $0\le \phi\le \phi_0$ with
\begin{eqnarray}\label{NBCtheta0}
\phi_0=\int_{F_{\text{min}}}^{\infty} \frac{F dF}{\bar{f}(F) \sqrt{\coth ^2\rho \ \bar{f}(F)-F^2}}. 
\end{eqnarray}
See Fig.\ref{Figureconeangle} for example, where $\bar{r}_h=1$, $\phi_0=\pi/2$ and  (\ref{NBCthetasolution})  covers one quarter of the angle range. For large $\rho\to \infty$, we also have $\phi_0\to \pi/2$. However, this is not the case for general $\rho$ and $\bar{r}_h$. 
To have a well-defined period, we require that 
\begin{eqnarray}\label{goodperiod} 
\frac{\pi q}{\phi_0}=\text{Integers},
\end{eqnarray}
which yields some constraints on $\rho$ and $\bar{r}_h$. In other words, the tensions of the branes $Q$ and $E$ are not independent for the cone holography with NBC.

Now let us turn to study the gravitational action. Following the approach of sect. 2.2 and using (\ref{NBCsect41}), we get 
\begin{eqnarray}\label{AdSCaction2NBCI}
I_{\text{AdSC}_{d+1}}
&=&\frac{1}{16\pi G_N}\int_0^{2\pi q} d\phi\int_{\bar{r}_h}^{\bar{r}_0} d\bar{r} \bar{r}^{d-1} \sqrt{|\bar{h}|}\left(\frac{R_{\bar{h}}+(d-1)(d-2)}{\bar{r}^2}-2d\right)\nonumber\\
&&+\frac{1}{8\pi G_N}\int_0^{2\pi q} d\phi\int_{\bar{Q}} \sqrt{|\bar{h}|} F(\phi)^{d-1} \sqrt{\frac{F'(\phi)^2}{\bar{f}(F(\phi))}+\bar{f}(F(\phi))} \tanh\rho \nonumber\\
&=& \frac{1}{16\pi G_N} \int_0^{2\pi q} d\phi \frac{F(\phi)^{d-2}-\bar{r}_h^{d-2}}{d-2} \int_{\bar{Q}} \sqrt{|\bar{h}|} \Big{(} R_{\bar{h}} +(d-2)(d-3) \Big{)} \nonumber\\
&=& \frac{1}{16\pi G^{(d-1)}_N}\int_{\bar{Q}} \sqrt{|\bar{h}|} \Big{(} R_{\bar{h}} +(d-2)(d-3) \Big{)}\ ,
\end{eqnarray}
which is proportional to the gravitational action of $\text{AdS}_{d-1}$.  Note that the above derivations applies to general $d$. For our propose, we focus on $d=4$ in this section. 

Unlike the case of MBC (\ref{AdSCaction2}), the effective Newton's constant is divergent
\begin{eqnarray}\label{NewtonconstantNBCI0}
\frac{1}{ G_N^{(3)}}=\frac{1}{ G_N} \int_0^{2\pi q} d\phi \frac{F(\phi)^{2}-\bar{r}_h^{2}}{2},
\end{eqnarray} 
where we have set $d=4$ above. 
That is because $F(\phi)$ could be infinite at some angles. See (\ref{Qcodim3NBCm1}) for example. To get finite results, we need to regularize the effective Newton's constant (\ref{NewtonconstantNBCI0}).  It is more convenient to consider the integral of $F$ instead of the integral of $\phi$ for (\ref{NewtonconstantNBCI0}). Using (\ref{NBCsect41},\ref{goodperiod}), we have
\begin{eqnarray}\label{NewtonconstantNBCI1}
\frac{1}{ G_N^{(3)}}=\frac{1}{ G_N} \frac{2\pi q}{\phi_0}\int_{F_{\text{min}}}^{F_{\infty}} \frac{F dF}{\bar{f}(F) \sqrt{\coth ^2\rho \ \bar{f}(F)-F^2}} \frac{F^{2}-\bar{r}_h^{2}}{2},
\end{eqnarray} 
where we set $F_{\infty} \to \infty$ at the end of regularization. Expanding the above integral element in powers of large $F$, we get
\begin{eqnarray}\label{NewtonconstantNBCI2}
\frac{1}{ G_N^{(3)}}&=&\frac{1}{ G_N} \frac{2\pi q}{\phi_0}\int_{F_{\text{min}}}^{F_{\infty}} \left(\frac{\sinh (\rho )}{2}+O(\frac{1}{F^2})\right) dF\nonumber\\
&=& \frac{1}{ G_N} \frac{2\pi q}{\phi_0} \left(\frac{\sinh (\rho )}{2} F_{\infty}+...\right)
\end{eqnarray} 
where $...$ denotes finite terms. One natural regularization is that we just drop the above divergent term and define the renormalized Newton's constant by
\begin{eqnarray}\label{regularizedGN}
\frac{1}{ G_{N\ \text{ren}}^{(3)}}= \frac{1}{ G_N} \frac{2\pi q}{\phi_0} \left(-\frac{\sinh (\rho )}{2} F_{\infty}+\int_{F_{\text{min}}}^{F_{\infty}} \frac{(F^{2}-\bar{r}_h^{2})F dF}{2\bar{f}(F) \sqrt{\coth ^2\rho \ \bar{f}(F)-F^2}}\right).
\end{eqnarray} 
Recall the fact that the induced metric on the brane at a finite place is dynamical, while the induced metric on the brane at infinity (such as the AdS boundary) is non-dynamical. The above regularization is just removing the contributions from the non-dynamical (infinite) regions of the brane. 
Note also that
\begin{eqnarray}\label{regularizedGNbyAdS}
\frac{\sinh (\rho )}{2} F_{\infty}=\lim_{\bar{r}_h\to 1} \int_{F_{\text{min}}}^{F_{\infty}} \frac{(F^{2}-\bar{r}_h^{2})F dF}{2\bar{f}(F) \sqrt{\coth ^2\rho \ \bar{f}(F)-F^2}},
\end{eqnarray} 
is the value for a pure AdS space. Thus (\ref{regularizedGN}) is just the usual regularization by subtracting an AdS background.  

\begin{figure}[t]
\centering
\includegraphics[width=12cm]{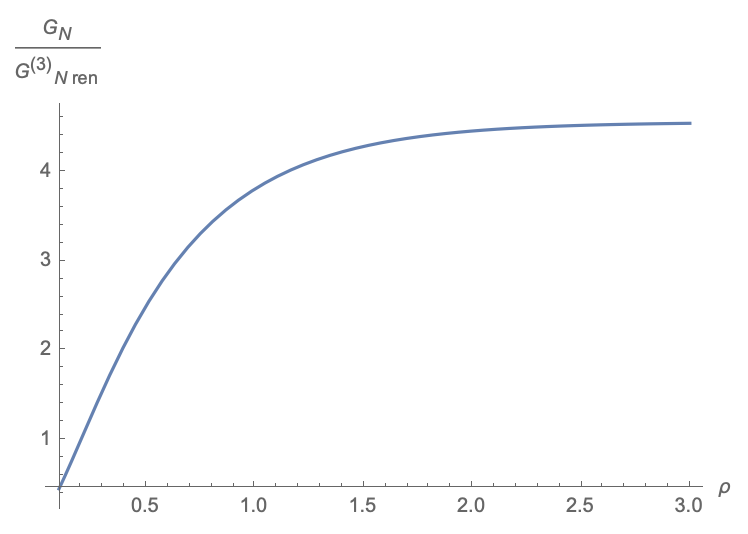}
\caption{The renormalized Newton's constant (\ref{regularizedGN}) increases with $\rho$, which is consistent with c-theorem. Without loss of generality, we have set $q=3$. }
\label{FigureNewtonconstant}
\end{figure}

Let us take the cases with $\phi_0=\pi/2$ (\ref{NBCtheta0}) to illustrate the above formula. Equivalently, we consider the cases $\bar{r}_h=1$ or $\rho\to \infty$. By direct calculations, we find that (\ref{regularizedGN}) vanishes for $\bar{r}_h=1$, which means that the dual CFT has zero central charge $c=3/(2G_{N\ \text{ren}}^{(3)})=0$. To have a non-trivial CFT duality, we require that $\bar{r}_h\ne 1$ for this model. As for  $\rho\to \infty$, we derive \footnote{Please see appendix B for the derivations. }
\begin{eqnarray}\label{regularizedGNlargerho}
\lim_{\rho\to \infty}\frac{1}{ G_{N\ \text{ren}}^{(3)}}= \frac{1}{ G_N} 4q \left(1-\bar{r}_h^2\right) \left[ \frac{ \pi }{8}  \left(\bar{r}_h^2+2\right)  -\frac{3\pi }{16}  \bar{r}_h^4 \left(5 \bar{r}_h^2-1\right) e^{-2\rho}+O(e^{-4\rho})\right],
\end{eqnarray} 
where $\frac{1}{\sqrt{2}}\le \bar{r}_h=\frac{1+\sqrt{1+8 q^2}}{4 q} < 1$ and we have $\partial_{\rho} c>0$ from (\ref{regularizedGNlargerho}).  Recall that $\rho_{UV}>\rho_{IR}$ and $c=3/(2G_{N\ \text{ren}}^{(3)})$, we notice that the c-theorem $c_{UV}\ge c_{IR}$ is obeyed for large $\rho$. In fact, one can check that the right hand side of (\ref{regularizedGN}) is a positive and monotonically increasing function of $\rho$. See Fig.\ref{FigureNewtonconstant} for example. Note that the $\rho$ satisfing the constraint (\ref{goodperiod}) are actually some discrete points in the blue line of Fig.\ref{FigureNewtonconstant}. Since the continuous $\rho$ is consistent with c-theorem, so does the discrete $\rho$. Thus the cone holography with NBC obeys the c-theorem. 

\begin{figure}[t]
\centering
\includegraphics[width=12cm]{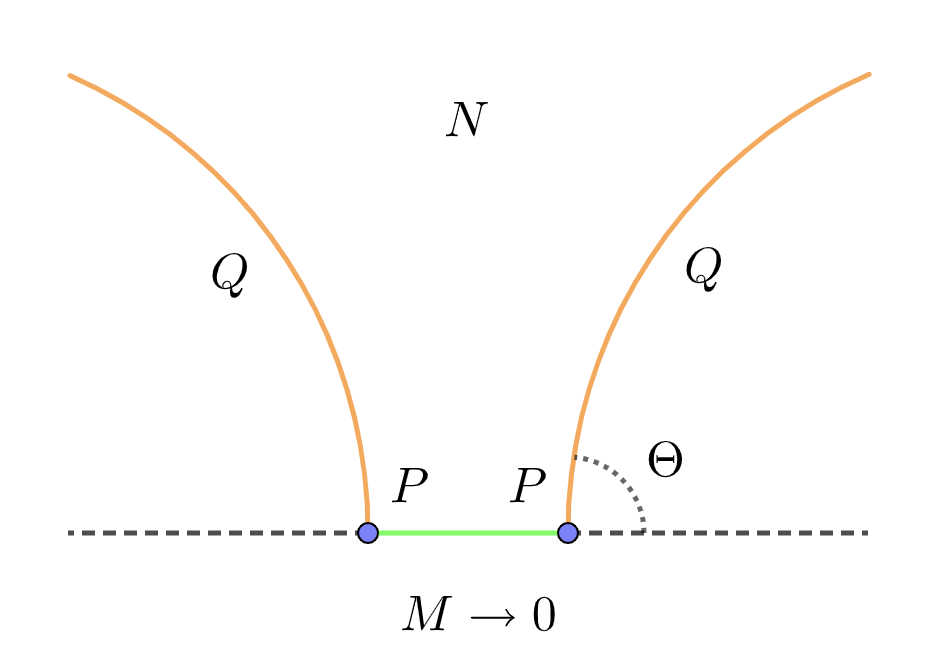}
\caption{Cone holography from AdS/BCFT. Taking the limit $M\to 0$, the above figure becomes the region near $\phi=0, \pi$ and $\bar{z}=1/\bar{r} \sim 0$ of Figure 5 for cone holography with NBC.}
\label{Figurerenormalization}
\end{figure}

In fact, the regularized Newton's constant (\ref{regularizedGN}) can be derived from the holographic renormalization of AdS/BCFT \cite{Chu:2017aab}.  Recall that the divergence of (\ref{NewtonconstantNBCI0},\ref{NewtonconstantNBCI1}) comes from the region near $\bar{r}=1/\bar{z}=F_{\infty}$ or equivalently, $\phi = 0, 2\phi_0, 4\phi_0,...$ .  As shown in Fig.\ref{Figurerenormalization}, these regions can be regarded as a special limit $M\to 0$ from AdS/BCFT. 

The renormalized action of $\text{AdS}_5/\text{BCFT}_4$ can be found in  \cite{Chu:2017aab}, where suitable counterterms are added on $M$ and $P$. See Fig.\ref{Figurerenormalization} for a sketch of  $M$ and $P$.  Taking the limit $M\to 0$, only the counterterms on $P$ survive for the cone holography with NBC \footnote{Note that the counterterms on $P$ also inculde extrinsic curvatures. Since they do not contribute to the present case, we do not list them for simplicity.}
\begin{eqnarray}\label{countertermsIP}
  I_P=\frac{2\pi q}{\phi_0} \frac{1}{8\pi G_N}\int_P \sqrt{|g_P|} (\Theta-\Theta_0-\frac{1}{4} \sinh\rho \ R_P),
\end{eqnarray}
where $P$ is located at $\bar{r}=F_{\infty}$, $\frac{2\pi q}{\phi_0} $ denotes the total numbers of $P$, $\Theta$ is the supplementary angle between $Q$ and $M$,  $\Theta_0=\Theta(\bar{r}\to \infty)$, $g_P$ denotes and metric and $R_P$ is Ricci scalar on $P$.  See Fig.\ref{Figurerenormalization} for example. Note that the first term of (\ref{countertermsIP}) is the famous Hayward term \cite{Hayward:1993my,Brill:1994mb}, which is added for a well-defined variation of the action. 

From  (\ref{metricBHcodim3}) with $d=4$ and (\ref{NBCsect41}), we can derive
\begin{eqnarray}\label{countertermsIP1}
  I_P= \frac{1}{ 16\pi G_N} \frac{2\pi q}{\phi_0} \left(\frac{-\sinh (\rho )}{2} F_{\infty}+O(\frac{1}{F_{\infty}}) \right) \int_{\bar{Q}} \sqrt{|\bar{h}|} ( R_{\bar{h}} +2 ).
\end{eqnarray}
Adding the above $I_P$ to the cone action (\ref{AdSCaction2NBCI}) and using (\ref{NewtonconstantNBCI2}), we finally obtain 
\begin{eqnarray}\label{coneholographyNBCItotalaction}
 I_{\text{AdSC}_5} + I_P= \frac{1}{ 16\pi G^{(3)}_{N \ \text{ren}}} \int_{\bar{Q}} \sqrt{|\bar{h}|} ( R_{\bar{h}} +2 )=I_{\text{AdS}_3},
\end{eqnarray}
 where $G^{(3)}_{N \ \text{reg}}$ exactly agrees with (\ref{regularizedGN}).  Now we finish the proof of the equivalence between the cone holography with NBC and AdS/CFT. We also verify that the renormalized Newton's constant is indeed given by (\ref{regularizedGN}) by applying the holographic renormalization.  Following approaches of sect.3, we can derive holographic Weyl anomaly , Entanglement/R\' enyi entropy, and correlation functions for the cone holography with NBC. 
 
 \section{Massive modes of cone holography}
 
 In the above sections, we focus on the class of solutions (\ref{metricMBCgeneral}), where the effective gravity on the brane is Einstein gravity and thus is massless. In this section, we investigate more general solutions to cone holography and find that there are infinite towers of massive modes of gravitons on the brane. As a result, the cone holography is different from AdS/CFT with Einstein gravity generally. In some range of parameters such as small brane tensions $T\ll 1$, the massive modes are suppressed and Einstein gravity is a good approximate on the brane.
 
 For simplicity, we focus on MBC in this section. We take the following ansatz of the perturbation metric
 \begin{eqnarray}\label{perturbationmetricI}
ds^2=dr^2+\sinh^2(r) d\Omega_{n-2}^2+\cosh^2 (r) \left( \bar{h}^{(0)}_{ij}(y) + H(r) \bar{h}^{(1)}_{ij}(y)  \right)dy^i dy^j,
\end{eqnarray}
where $\bar{h}^{(0)}_{ij}(y)$ is an AdS metric and $\bar{h}^{(1)}_{ij}(y)$ denote the metric perturbations on the brane $E$. Note that the induced metric on $E$ is proportional to the AdS sector of the induced metric on $Q$.  
In the language of bulk metric perturbations,  we have
 \begin{eqnarray}\label{bulkmetricperturbations}
\delta g_{rA}=0,\ \delta g_{aA}=0,\ \delta g_{ij}=\cosh^2 (r)  H(r) \bar{h}^{(1)}_{ij}(y).
\end{eqnarray}
For simplicity we do not consider the angle dependence of perturbations.

We choose the following gauge 
 \begin{eqnarray}\label{gijgauge}
\nabla^A \delta g_{AB}=0,\ \ \  g^{AB}\delta g_{AB}=0,
\end{eqnarray}
which yields
 \begin{eqnarray}\label{hij1gauge}
D^i \bar{h}^{(1)}_{ij}=0,\ \ \  \bar{h}^{(0)ij}\bar{h}^{(1)}_{ij}=0,
\end{eqnarray}
where $\nabla^A$ and $D^i$ are the covariant derivatives with respect to $g_{AB}$ and $\bar{h}^{(0)}_{ij}$, respectively.  
In the gauge (\ref{gijgauge}), Einstein equations become
 \begin{eqnarray}\label{EOMgABgauge}
 \left(\nabla_C \nabla^C+2\right) \delta g_{AB}=0.
\end{eqnarray}
Substituting (\ref{bulkmetricperturbations}) together with (\ref{hij1gauge})  into (\ref{EOMgABgauge}) and separating variables, we obtain
 \begin{eqnarray}\label{EOMMBCmassivehij}
&& \left(D_i D^i+2-\hat{m}_k^2\right) \bar{h}^{(1)}_{ij}(y)=0,\\
&& \sinh (2 r) H''(r)+(d \cosh (2 r)-d+2 n-4)H'(r) +2 \hat{m}_k^2  \tanh (r) H(r)=0, \label{EOMMBCmassiveH}
\end{eqnarray}
where $\hat{m}_k$ denotes the mass of gravitons on the brane, which will be determined later. Solving (\ref{EOMMBCmassiveH}), we get
 \begin{eqnarray}\label{EOMMBCmassiveHsolution}
H(r)=
\begin{cases}
c_1 \, _2F_1\left(a_1,a_2;\frac{n-1}{2};\tanh ^2(r)\right)+c_2 G_{2,2}^{2,0}\left(\tanh ^2(r)|
\begin{array}{c}
 a_1+\frac{d}{2}, a_2+\frac{d}{2} \\
 0,0 \\
\end{array}
\right),&n=3,\\
c_1 \, _2F_1\left(a_1,a_2;\frac{n-1}{2};\tanh ^2(r)\right)+c_2 \tanh ^{3-n}(r) \, _2F_1\left(a_3,a_4;\frac{5-n}{2};\tanh ^2(r)\right),&n>3,
\end{cases}
\end{eqnarray}
where $_2F_1$ is the hypergeometric function, $G_{2,2}^{2,0}$ is the Meijer G function, $c_1$ and $c_2$ are integral constants and $a_i$ are given by
 \begin{eqnarray}\label{aibia1}
&&a_1=\frac{1}{4} \left(n-d-1-\sqrt{(d-n+1)^2+4 \hat{m}_k^2}\right),\\ \label{aibia2}
&&a_2=\frac{1}{4} \left(n-d-1+\sqrt{(d-n+1)^2+4 \hat{m}_k^2}\right),\\ \label{aibia3}
&&a_3=\frac{1}{4} \left(5-d-n-\sqrt{(d-n+1)^2+4 \hat{m}_k^2}\right),\\ \label{aibia4}
&&a_4=\frac{1}{4} \left(5-d-n+\sqrt{(d-n+1)^2+4 \hat{m}_k^2}\right).
\end{eqnarray}
Near the brane $E$ ($r=0$), (\ref{EOMMBCmassiveHsolution}) behaves as
 \begin{eqnarray}\label{EOMMBCmassiveHseries}
H(r)\sim
\begin{cases}
c_2 \ln r+...,& n=3,\\
c_2\ \frac{1 }{r^{n-3}}+...,& n>3,
\end{cases}
\end{eqnarray}
where `...' denote higher order terms in $r$. 

We impose the natural boundary condition on the brane $E$ ($r=0$), which means the perturbation on $E$ is finite,
 \begin{eqnarray}\label{naturalBCHI}
H(r)|_{r=0} \ \text{is finite}.
\end{eqnarray}
From (\ref{EOMMBCmassiveHseries}, \ref{naturalBCHI}), we get
 \begin{eqnarray}\label{c2I}
c_2=0.
\end{eqnarray}
We impose MBC (\ref{MBC}) on the end-of-world brane $Q$ ($r=\rho$), which yields
 \begin{eqnarray}\label{MBCHI}
H'(r)|_{r=\rho} =0.
\end{eqnarray}
Substituting (\ref{EOMMBCmassiveHsolution},\ref{c2I}) into (\ref{MBCHI}), we derive
 \begin{eqnarray}\label{BCformass}
\hat{m}_k^2 \frac{c_1 }{1-n}  \tanh (\rho ) \text{sech}^2(\rho ) \, _2F_1\left(1+a_1,1+a_2;\frac{n+1}{2};\tanh ^2(\rho )\right)=0.
\end{eqnarray}
Recall that $a_i$ are given by (\ref{aibia1},\ref{aibia2}) which depend on $\hat{m}_k^2$. The above equation imposes a constraint on the possible mass $\hat{m}_k$.  Clearly, the massless mode with $\hat{m}_0^2=0$ is always a solution to (\ref{BCformass}). 

To get more understandings of the spectrum, let us study some special cases. In the large $\rho$ limit, (\ref{BCformass}) can be approximately by
 \begin{eqnarray}\label{BCformasslarge}
&&\, _2F_1\left(1+a_1,1+a_2;\frac{n+1}{2};1\right)\nonumber\\
&&=\frac{\Gamma \left(\frac{d}{2}-1\right) \Gamma \left(\frac{n+1}{2}\right)}{\Gamma \left(\frac{d+n-1-\sqrt{4 \hat{m}_k^2+(d-n+1)^2}}{4} \right) \Gamma \left(\frac{d+n-1+\sqrt{4 \hat{m}_k^2+(d-n+1)^2}}{4}\right)}\approx 0,
\end{eqnarray}
which has the roots
 \begin{eqnarray}\label{masslargerho}
\hat{m}^2_k \approx (2 k+d-2) (2 k+n-3), \ \ k\ge 1 ,\ \ \ \text{for large} \ \rho.
\end{eqnarray}
Recall that $\rho$ is related to the tension of the end-of-world brane $Q$, i,e, $T=(d-1) \tanh(\rho)$. Note that the mass $\hat{m}_k$ defined in (\ref{EOMMBCmassivehij}) is with respective to the induced metric $\bar{h}^{(0)}_{ij}$ on $E$. We are interested in the effective mass on $Q$ instead of $E$. That is because, according to the brane world holography, the gravity is approximately localized on the end-of-world brane $Q$. The mass on $Q$ is defined by
 \begin{eqnarray}\label{massonQdefinition}
(\Box_Q+\frac{2}{L^2_Q}-\hat{m}^2_{Q \ k}) \bar{h}^{(1)}_{ij}(y)=(\frac{D^iD_i+2}{\cosh^2(\rho)}-\hat{m}^2_{Q \ k}) \bar{h}^{(1)}_{ij}(y)=0,
\end{eqnarray}
where $\Box_Q$ and $L_Q=\cosh(\rho)$ are the D'Alembert operator and the AdS radius on $Q$, respectively. Note that the induce metric on $Q$ is $\cosh^2(\rho)\bar{h}^{(0)}_{ij}$. As a result, we have $\Box_Q=D^iD_i/\cosh^2(\rho)$.
Comparing (\ref{massonQdefinition}) with (\ref{EOMMBCmassivehij}), we read off 
 \begin{eqnarray}\label{relationofmass}
\hat{m}^2_{Q \ k}=\frac{\hat{m}^2_{\ k}}{\cosh^2(\rho)}.
\end{eqnarray}
From (\ref{masslargerho}) and (\ref{relationofmass}), we find that the mass on $Q$ becomes continuous in the large $\rho$ limit
 \begin{eqnarray}\label{massonQlarge}
\lim_{\rho\to\infty}\Delta \hat{m}_{Q \ k}\sim \frac{1}{\cosh(\rho)} \to 0.
\end{eqnarray}
Since the range of $r$ is infinite in the large $\rho$ limit., i.e., $0\le r \le \rho\to \infty$, it is natural that the spectrum becomes continuous in this case.

Let us go on to study the limit with small $\rho$.  From (\ref{relationofmass}), we have $\hat{m}_{Q \ k}\approx \hat{m}_{ k}$ at small $\rho$. Thus we do not distinguish them below. Recall that $0\le r \le \rho$, small $\rho$ also means small $r$. For small $r$,  EOM (\ref{EOMMBCmassiveH}) becomes
 \begin{eqnarray}
 2r H''(r)+(2 n-4)H'(r) +2 \hat{m}_k^2\  r H(r)=0. \label{EOMMBCmassiveHsmallrho}
\end{eqnarray}
Remarkably, the above equation is independent of the spacetime dimension $d$.  
Solving the above equation together with the natural boundary condition (\ref{naturalBCHI}), we get
 \begin{eqnarray} \label{Hforsmallr}
H(r)=c_1 r^{\frac{3}{2}-\frac{n}{2}} J_{\frac{n-3}{2}}(|\hat{m}_k| r),
\end{eqnarray}
where $J_v$ denotes the Bessel function of the first kind.
Now imposing the MBC (\ref{MBCHI}), we obtain the constraint of $\hat{m}_k $ for small $\rho$,
 \begin{eqnarray} \label{Hforsmallr}
c_1 \rho ^{\frac{3}{2}-\frac{n}{2}}\ |\hat{m}_k|\  J_{\frac{n-1}{2}}(|\hat{m}_k| \rho )=0.
\end{eqnarray}
Again, the massless mode $\hat{m}_0=0$ is a solution.  In general, the masses $|\hat{m}_k|$ are given by the roots of Bessel function $J_{\frac{n-1}{2}}$ divide by $\rho$. Let us list the first few terms of $|\hat{m}_k|$ \begin{eqnarray} \label{massforsmallrhon3}
&&|\hat{m}_k|= 0,\ \frac{3.83}{\rho} ,\ \frac{7.02}{\rho} ,\ \frac{10.17}{\rho} ,\ \frac{13.32}{\rho} ,...\ \ \text{for n=3},\\ \label{massforsmallrhon4}
&&|\hat{m}_k|= 0,\ \frac{4.49}{\rho} ,\ \frac{7.73}{\rho} ,\ \frac{10.90}{\rho} ,\ \frac{14.07}{\rho} ,...\ \ \text{for n=4}.
\end{eqnarray}
The asymptotic expression of (\ref{Hforsmallr}) at large $|\hat{m}_k| \rho$ is given by
 \begin{eqnarray} \label{Hforsmallrlargek}
c_1 \rho ^{1-\frac{n}{2}}\ \sqrt{\frac{2 \hat{m}_k }{\pi  }} \cos(|\hat{m}_k| \rho-\frac{n\pi}{4})\approx 0,
\end{eqnarray}
from which we derive
 \begin{eqnarray}\label{masssmallrho}
|\hat{m}_k| \approx \frac{\pi  \left(4 k-4n_1+n-2\right)}{4 \rho }, \ \ k\gg 1 ,\ \ \ \text{for small} \ \rho,
\end{eqnarray}
where $n_1$ is an integer which we fine turn in order to make (\ref{masssmallrho}) to be the k-th root of (\ref{Hforsmallr}) for large $k$. For example, we have $n_1=0$ for $n=3$ and $n=4$. Interestingly, (\ref{masssmallrho}) agrees well with (\ref{massforsmallrhon3},\ref{massforsmallrhon4}) even for small $k$. 

 \begin{figure}[t]
\centering
\includegraphics[width=10cm]{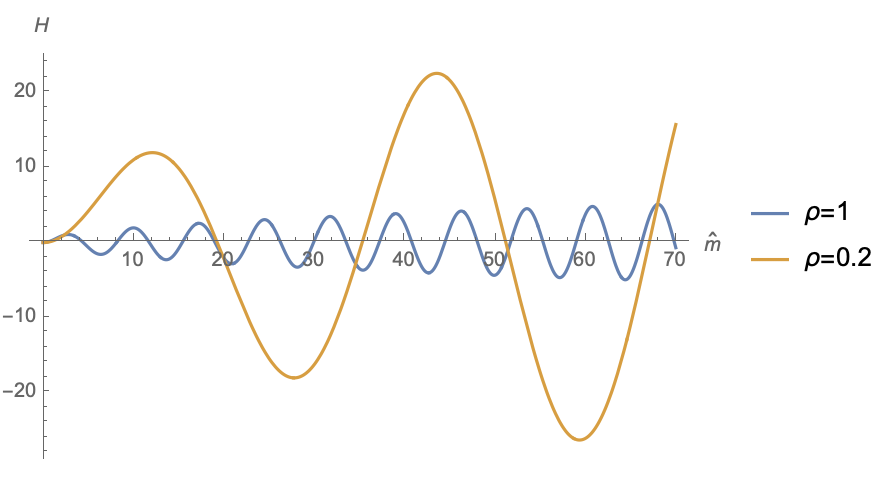}
\includegraphics[width=10cm]{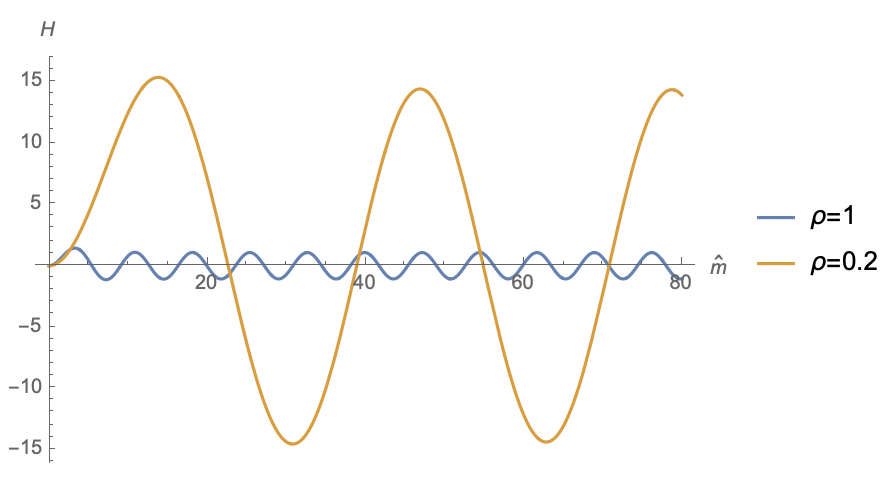}
\caption{Mass spectrum of cone holography, where the masses correspond to the intersections of the curve $H'(\rho)$ and the $\hat{m}$-axis. The above figure is for $d=4$ and $n=3$, and the below figure is for $d=5$ and $n=4$.  The blue curve is for $\rho=1$ and the yellow curve is for $\rho=0.2$. The larger the tension $\rho$ is, the more continuous the mass spectrum is.}
\label{mass4d3n}
\end{figure}

Some comments are in order. First, it is remarkable that, at small $\rho$, $\hat{m}_k$ is independent of the dimension of spacetime, i.e., $d$.  Second,  $\hat{m}_k$ with $m\ge 1$ is inversely proportional to $\rho$ and thus is quite large at small $\rho$. As a result, the massive modes of gravitons are suppressed and the low-energy effective theory on the brane is Einstein gravity approximately. It is quite similar to the usual KK mechanism. Consider a space with a small circle $R_4\times S_1$, the masses of massive KK modes are quite large, i.e., $\hat{m}_k \sim k/r_0 \to \infty$, when the circle radius is small, i.e., $r_0\to 0$.  At low energy, only the massless mode is excited and the massive modes are frozen. Third, the cone holography with $0\le r \le \rho$ is quite different from the usual brane world holography with $-\infty\le r \le \rho$. As a result, the mass spectrums of the two theories are quite different. For example, the masses of  brane world holography are not inversely proportional to $\rho$ at small $\rho$. 

To summarize, we find that, as it is expected, there are infinite massive KK modes on the brane.  As a result, in general, the cone holography is different from AdS/CFT with Einstein gravity.  One can argue that the cone holography is equivalent to AdS/CFT with infinite towers of massive gravity on the brane.  However, the simpler and better way to study the holographic dual of edge modes is the Einstein gravity is the bulk cone rather than 
the infinite towers of massive gravity on the brane.  For the large brane tension $\rho \to \infty$, the mass spectrum is continuous. While for the small brane tension $\rho\to 0$, the mass spectrum is discrete and the masses of massive modes become infinite.  At low energy, only the massless mode is excited. As a result, when $\rho$ is small, the low-energy effective theory on the brane is Einstein gravity.  To end this section, let us draw some figures for the mass spectrum of cone holography. See Fig.\ref{mass4d3n} for example, where the intersections of the curve and the $\hat{m}$-axis denote the allowed masses on the brane.

\section{Conclusions and Discussions}

In this paper, we propose a novel codim-n holography, called cone holography, which conjectures that a gravity theory in $(d+1)$-dimensional conical spacetime is dual to a CFT on the $(d+1-n)$-dimensional defects. The cone holography can be derived by taking the suitable zero-volume limit of AdS/dCFT and it can be regarded as a holographic dual of the edge modes on the defects. For one class of exact solutions, we prove the cone holography by showing that it is equivalent to AdS/CFT with Einstein gravity. The proof is valid at least in the classical level for gravity, or equivalently, in large N limit for CFTs.  We test cone holography by studying holographic Weyl anomaly, holographic R\'enyi entropy and correlation functions, and find good agreements with the results of CFTs. In particular, the c-theorem is obeyed by cone holography. These are strong supports for our proposal.  In addition to the mixed boundary condition, we also discuss the cone holography with Neumann boundary conditions.  We find that the end-of-world brane $Q$ intersects with the AdS boundary $M$ at some specific angles for NBC. As a result, the effective Newton's constant is divergent and needs to be regularized. By performing the holographic renormalization, we get a well-defined Newton's constant, which is consistent with the c-theorem.  Finally, we analyze the mass spectrum of cone holography and find that the larger the tension is, the more continuous the mass spectrum is. Due to the massive KK modes on the brane, in general, cone holography is different from AdS/CFT with Einstein gravity. When the tension $\rho$ is small, since the massive modes are frozen at low energy, the effective theory on the brane is Einstein gravity.  The cone holography is a generalization of wedge holography \cite{Akal:2020wfl,Bousso:2020kmy,Miao:2020oey}, and is closely related to brane-world holography \cite{Randall:1999ee,Randall:1999vf,Karch:2000ct}, AdS/BCFT \cite{Takayanagi:2011zk,Fujita:2011fp,Nozaki:2012qd,Miao:2018qkc,Miao:2017gyt,Chu:2017aab}, AdS/dCFT \cite{Jensen:2013lxa,DeWolfe:2001pq,Dong:2016fnf} and holographic Entanglement/R\'enyi entropy \cite{Dong:2016fnf,Hung:2011nu,Dong:2016wcf,Ryu:2006bv,Chu:2016tps}. Thus it is expected to have a wide ranges of applications. 

Many interesting problems remain to be investigated. We list some of them for examples. 

{\bf 1}. For the cone holography with NBC, we assume the embedding function of $Q$ to be $\bar{r}=F(\phi)$ and find that $Q$ tends to infinity at some angles, i.e., $F(\phi_i )\to \infty$. It is interesting to consider more general embedding functions and see if the end-of-world brane $Q$ could be located at finite place. 

{\bf 2}. There are many different kinds of warped embeddings between Einstein manifolds \cite{Yang:2010mca}. In this paper, we discuss only one of them and find that the corresponding cone holography is equivalent to AdS/CFT with Einstein gravity.  It is interesting to study other kinds of embeddings in the framework of cone holography. 


{\bf 3}. Find more general solutions different from (\ref{metricMBCgeneral}) of cone holography. Similar to the case of wedge holography \cite{Miao:2020oey}, these solutions are expected to reproduce more general Weyl anomaly, such as the second and third terms of (\ref{Weylanomaly4ddefect2}).

{\bf 4}. In this paper, we mainly focus on vacuum Einstein gravity.  We discuss briefly the scalar fields for correlation functions.  It is interesting to study more general matter fields such as Maxwell's fields, which play an important role in AdS/CMT.

{\bf 5}. Generalize cone holography to higher derivative gravity. It is interesting that, unlike the wedge holography, it is easier to find non-trivial solutions (non-AdS) to the cone holography for higher derivative gravity.  

{\bf 6}. In this paper, we focus on the classical limit of gravity. It is interesting to study the quantum corrections and see if the equivalence between the cone holography and AdS/CFT still hold. 

{\bf 7}. Apply cone holography to discuss the information paradox such as island and the Page curve of Hawking radiations.  

{\bf 8}. Find other interesting applications for cone holography. 

We hope these interesting problems could be addressed in the future.

\section*{Acknowledgements}

We thank J. Ren and P. J. Hu for valuable discussions. This work is supported by NSFC grant (No. 11905297) and Guangdong Basic and Applied Basic Research Foundation (No.2020A1515010900).

\appendix

\section{Some formulas}

Let us start with the following ansatz of metric
\begin{eqnarray}\label{appendixmetric}
ds^2=g_{AB}dX^AdX^B=dr^2+f(r) \gamma_{ab}(x)dx^adx^b+ g(r) \bar{h}_{ij}(y)dy^i dy^j,
\end{eqnarray}
where $\gamma_{ab}$ and $ \bar{h}_{ij}$ are the metrics of Einstein manifolds.
\begin{eqnarray}\label{Einsteinmanifold}
R_{\gamma\ a b}=(n-3)  \gamma_{ab},\ \  R_{\bar{h}\ ij}=-(d+1-n)  \bar{h}_{ij}
\end{eqnarray}
From the above metric, we obtain non-zero affines and curvatures as follows
\begin{eqnarray}\label{affine1}
&&\Gamma^r_{ab}=-\frac{f'}{2f}g_{ab}, \ \ \Gamma^a_{rb}=\frac{f'}{2f}\delta^a_b,\ \ \Gamma^a_{bc}= \Gamma_{\gamma}{}^a_{bc},\\ \label{affine2}
&&\Gamma^r_{ij}=-\frac{g'}{2g}g_{ij}, \ \ \Gamma^i_{rj}=\frac{g'}{2g}\delta^i_j,\ \ \Gamma^i_{jk}= \Gamma_{\bar{h}}{}^i_{jk},
\end{eqnarray}
\begin{eqnarray}\label{curvature1}
R^r_{\ arb}=(-\frac{1}{2}\frac{f''}{f}+\frac{1}{4}\frac{f'^2}{f^2})g_{ab}, \ \  R^a_{\ bcd}=R_{\gamma}{}^a_{\ bcd}-\frac{1}{4} \frac{f'^2}{f^2}(\delta^a_c g_{bd}-\delta^a_d g_{bc}),
\end{eqnarray}
\begin{eqnarray}\label{curvature2}
R^r_{\ irj}=(-\frac{1}{2}\frac{g''}{g}+\frac{1}{4}\frac{g'^2}{g^2})g_{ij}, \ \  R^i_{\ jkl}=R_{\bar{h}}{}^i_{\ jkl}-\frac{1}{4} \frac{g'^2}{g^2} \ (\delta^i_k g_{jl}-\delta^i_l g_{jk}),
\end{eqnarray}
\begin{eqnarray}\label{curvature3}
R^a_{\ ibj}=-\frac{1}{4}  \frac{f'}{f}\frac{g'}{g} \ \delta^a_b g_{ij},
\end{eqnarray}
\begin{eqnarray}\label{curvature4}
R_{rr}=(d-n+2) \left(\frac{\left(g'\right)^2}{4 g^2}-\frac{g''}{2 g}\right)+(n-2) \left(\frac{\left(f'\right)^2}{4 f^2}-\frac{f''}{2 f}\right),
\end{eqnarray}
\begin{eqnarray}\label{curvature5}
R_{ab}= R_{\gamma\ ab}+\left(\frac{\left(f'\right)^2}{4 f^2}-\frac{(d-n+2) f' g'}{4 f g}-\frac{f''}{2 f}-\frac{(n-3) \left(f'\right)^2}{4 f^2}\right) g_{ab} 
\end{eqnarray}
\begin{eqnarray}\label{curvature6}
R_{ij}= R_{\bar{h}\ ij}+ \left(\frac{\left(g'\right)^2}{4 g^2}-\frac{(d-n+1) \left(g'\right)^2}{4 g^2}-\frac{(n-2) f' g'}{4 f g}-\frac{g''}{2 g}\right) g_{ij}
\end{eqnarray}
\begin{eqnarray}\label{curvature7}
R&=&\frac{R_{\gamma} }{f}+\frac{R_{\bar{h}} }{g}-\frac{(n-2) f' \left(2 f (d-n+2) g'+g (n-5) f'\right)}{4 f^2 g}\nonumber\\
&&-\frac{4 g \left(f (d-n+2) g''+g (n-2) f''\right)+f (d-n-1) (d-n+2) \left(g'\right)^2}{4 f g^2}
\end{eqnarray}
where $g_{ab}=f(r) \gamma_{ij}$, $g_{ij}=g(r) \bar{h}_{ij}$, $(\ )_{\gamma}$ and $(\ )_{\bar{h}}$ denotes the quantities defined by the metrics $\gamma_{ij}$ and $\bar{h}_{ij}$, respectively. 

Applying Einstein equations $R_{AB}=-d g_{AB}$ together with (\ref{Einsteinmanifold}), we obtain EOMs of $f(r)$ and $g(r)$ as
\begin{eqnarray}\label{EOMfg1}
&&\frac{(d-n+2) \left(\left(g'\right)^2-2 g g''\right)}{4 g^2}+d+\frac{(n-2) \left(\left(f'\right)^2-2 f f''\right)}{4 f^2}=0,\\ \label{EOMfg2}
&&\frac{(-d+n-2) f' g'+2 g \left(-f''+2 n-6\right)}{4 f g}+d-\frac{(n-4) \left(f'\right)^2}{4 f^2}=0, \\ \label{EOMfg3}
&&\left(-2 g \left(2 d+g''-2 n+2\right)+4 d g^2+(n-d) \left(g'\right)^2\right)+\frac{ (2-n) gf' g'}{f}=0.
\end{eqnarray}
There are two kinds of exact solutions to the above equations. The first kind is 
\begin{eqnarray}\label{appsolutions1}
f(r)=\sinh^2r,\ \  g(r)=\cosh^2r,
\end{eqnarray}
which corresponds to an asymptotically AdS space. The second kind of solution is given by
\begin{eqnarray}\label{appsolutions2}
f(r)=\frac{n-2}{d} \sinh^2(\frac{\sqrt{d} }{\sqrt{n-2}} r),\ \  g(r)=\frac{d-n+1}{d}.
\end{eqnarray}
Note that the metric (\ref{appendixmetric}) with $f(r)$ and $g(r)$ given by (\ref{appsolutions2}) is not an AdS metric. 

From (\ref{EOMfg1},\ref{EOMfg2},\ref{EOMfg3}), we can derive two independent equations
\begin{eqnarray}\label{EOMfg4}
g=\frac{f (d-n+1) (d-n+2) \left(f'\right)^2}{\left(f'\right)^2 \left((d-6) d f+f''+(n-6) (n-3)\right)+f f^{(3)} f'-2 f \left(d f-f''+n-3\right) \left(2 (d f+n-3)-f''\right)}\nonumber\\
\end{eqnarray}
and 
\begin{eqnarray}\label{EOMfg5}
&&16 d^2 f^4+(d (n-6)+4) \left(f'\right)^4-4 f \left(f'\right)^2 \left(2 \left(n^2-7 n+12\right)-(d-2) f''\right)\nonumber\\
&&+4 f^2 \left(d (d-3 n+10) \left(f'\right)^2\right)-8 d f^3 \left((d-n+4) f''-4 (n-3)\right)\nonumber\\
&&+4 f^2 \left(\left(-f''+2 n-6\right) \left((-d+n-3) f''+2 (n-3)\right)+(-d+n-2) f' f^{(3)} \right)=0.\nonumber\\
\end{eqnarray}
The 3rd order differential equation (\ref{EOMfg5}) can be solved Numerically.  Once we solve $f(r)$ from (\ref{EOMfg5}) , we can derive $g(r)$ by (\ref{EOMfg4}).   

To end this section, let us consider the perturbation solutions near the brane $E$, i.e., $r=0$. We take the following ansatzs,
\begin{eqnarray}\label{appansatzfg1}
&&f(r)=\frac{r^2}{\bar{m}^2}+c_1 r^4 +c_2 r^6+O(r^8),\\ \label{EOMfg8}
&&g(r)=d_1+d_2 r^2 +O(r^4).
\end{eqnarray}
Substituting the above ansatzs into Einstein equation (\ref{EOMfg2}), we get
\begin{eqnarray}\label{appansatzfg2}
\frac{\left(\bar{m}^2-1\right) (n-3)}{r^2}+O(r^0)=0,
\end{eqnarray}
which yields either $n=3$ or $\bar{m}=1$.   For $n=3$, solving EOMs (\ref{EOMfg1},\ref{EOMfg2},\ref{EOMfg3}), we derive 
\begin{eqnarray}\label{appansatzfgn3}
&&c_2=\frac{c_1^2 (17 d-8) \bar{m}^4-3 c_1 d (d+1) \bar{m}^2+d^2}{20 (d-1) \bar{m}^2},\nonumber\\
&&d_1=\frac{(d-2) (d-1)}{6 c_1 \bar{m}^2+(d-3) d},\nonumber\\
&&d_2=\frac{(d-2) \left(d-3 c_1 \bar{m}^2\right)}{6 c_1 m^2+(d-3) d},
\end{eqnarray}
where $\bar{m}$ and $c_1$ are free parameters. As for $n>3$, we must have $\bar{m}=1$. Solving EOMs (\ref{EOMfg1},\ref{EOMfg2},\ref{EOMfg3}) we obtain 
\begin{eqnarray}\label{appansatzfgnne3}
&&c_2=\frac{c_1^2 \left(d (11 n-16)-2 n^2+2 n+4\right)-3 c_1 d (d+n-2)+d^2}{5 (n+1) (d-n+2)},\nonumber\\
&&d_1=\frac{(d-n+1) (d-n+2)}{3 c_1 \left(n^2-3 n+2\right)+d (d-2 n+3)},\nonumber\\
&&d_2=\frac{(d-n+1) \left(d-3 c_1 (n-2)\right)}{3 c_1 \left(n^2-3 n+2\right)+d (d-2 n+3)},
\end{eqnarray}
where $c_1$ is a free parameter. Similarly, we can solve the perturbation solutions near the AdS boundary $M$, i.e., $r\to \infty$. One can check that $(\lim_{r\to \infty} \frac{f(r) }{g(r)}-1)$ can be non-zero for $n>3$. Thus the BC (\ref{conicalBCnne3}) chosen in sect. 2.1 is well-defined. 

\section{The integral}

In this appendix, we study the integral (\ref{regularizedGN}) carefully. Under the coordinate transformation 
\begin{eqnarray}\label{appcoordinatetransformation }
F= \frac{1}{2} \sqrt{\cosh (\rho ) \left(\sqrt{2} \left(2 R^2+1\right) \sqrt{8 \bar{r}_h^4-8 \bar{r}_h^2+\cosh (2 \rho )+1}+2 \cosh (\rho )\right)},
\end{eqnarray}
the integral (\ref{regularizedGN}) becomes
\begin{eqnarray}\label{appintegral}
&&\frac{1}{ G_{N\ \text{ren}}^{(3)}}= \frac{1}{ G_N} \frac{2\pi q}{\phi_0}\int_{0}^{\infty} dR \times \nonumber\\
&&\Big( \frac{\sinh (\rho ) \left(\cosh (\rho ) \left(2 \cosh (\rho )+\sqrt{2} \left(2 R^2+1\right) \sqrt{\cosh (2 \rho )+8 \bar{r}_h^4-8 \bar{r}_h^2+1}\right)\right)^{3/2}}{4 \sqrt{R^2+1} \left(\cosh (2 \rho )+\sqrt{2} \left(2 R^2+1\right) \cosh (\rho ) \sqrt{\cosh (2 \rho )+8 \bar{r}_h^4-8 \bar{r}_h^2+1}+4 \bar{r}_h^2-3\right)} \nonumber\\
&&-\frac{ 2^{3/4}}{4} \sinh (\rho ) \sqrt{\cosh (\rho )} \sqrt[4]{1+\cosh (2 \rho )+8 \bar{r}_h^4-8 \bar{r}_h^2} \Big).
\end{eqnarray} 
Expanding the above integral in powers of $e^{\rho}$, we have
\begin{eqnarray}\label{appintegral1}
\frac{1}{ G_{N\ \text{ren}}^{(3)}}&=& \frac{1}{ G_N} \frac{2\pi q}{\phi_0}\int_{0}^{\infty} dR \times \nonumber\\
&&\left(\frac{2-\bar{r}_h^4-\bar{r}_h^2}{4 \left(R^2+1\right)}+\frac{\left(\bar{r}_h^2-1\right) \left(R^2 \left(8 \bar{r}_h^6-6 \bar{r}_h^2+4\right)+7 \bar{r}_h^6-3 \bar{r}_h^4+6 \bar{r}_h^2-4\right)}{4 \left(R^2+1\right)^2} e^{-2\rho} +O(e^{-4\rho})\right)\nonumber\\
&=&\frac{1}{ G_N} 4q \left(1-\bar{r}_h^2\right) \left[ \frac{ \pi }{8}  \left(\bar{r}_h^2+2\right)  -\frac{3\pi }{16}  \bar{r}_h^4 \left(5 \bar{r}_h^2-1\right) e^{-2\rho}+O(e^{-4\rho})\right],
\end{eqnarray} 
which exactly agrees with (\ref{regularizedGNlargerho}). Note that we have used $\phi_0=\pi/2$ for $\rho\to \infty$ above. From (\ref{appintegral}), it is straightforward to numerically derive the relation between the renormalized Newton's constant and $\rho$. See Fig.\ref{FigureNewtonconstant} for example.

\end{document}